\documentclass[twocolumn,showpacs,prb]{revtex4}
\usepackage{graphicx}

\newcommand{\BEQ}{\begin{equation}}
\newcommand{\EEQ}{\end{equation}}
\newcommand{\BEA}{\begin{eqnarray}}
\newcommand{\EEA}{\end{eqnarray}}
\renewcommand{\d}{{\rm d }}
\newcommand{\p}{\partial}
\newcommand{\nn}{\nonumber }
\newcommand{\Tr}{{\rm Tr}}
\newcommand{\s}{{\sigma}}
\newcommand{\ext} {\mathop{\rm ext}}

\begin{document}

\title{Quenched Computation of the Complexity of the
Sherrington-Kirkpatrick Model.}  \author{A. Crisanti, L. Leuzzi,
G. Parisi, T. Rizzo} 
\affiliation{Department of Physics and INFM SMC
Center, University of Rome I, ``La Sapienza'', Piazzale A. Moro 2,
00185 Rome, Italy.}  
\date{\today} 
\begin{abstract}
The quenched computation of the complexity in the
 Sherrington-Kirkpatrick model is presented. A modified Full Replica
 Symmetry Breaking Ansatz is introduced in order to study the
 complexity dependence on the free energy. Such an Ansatz corresponds
 to require Becchi-Rouet-Stora-Tyutin supersymmetry.  The complexity
 computed this way is the Legendre transform of the free energy
 averaged over the quenched disorder. The stability analysis shows
 that this complexity is inconsistent at any free energy level but the
 equilibirum one.  The further problem of building a physically well
 defined solution not invariant under supersymmetry and predicting an
 extensive number of metastable states is also discussed.
\end{abstract}

\maketitle

\section*{Introduction}
\label{s:intro}

The frozen phase of mean field spin glass models displays, in the
thermodynamic limit, a very high number of stable and metastable
states.  Such a feature is the consequence of the disorder and the
frustration characterizing spin glasses and causing the onset of
many different configurations of spins minimizing the thermodynamic
potential, organized in the phase space in a rather complicated
way.  
In order to comprehend the structure of the landscape of the
thermodynamic potential below a given critical temperature a very
important theoretical tool is the {\em complexity}, else called, in
the framework of structural glasses, {\em configurational entropy},
i.e. the logarithm of the number of metastable states the system can
evolve to.
The organization of thermodynamic states
in complex systems is fundamental, e.g., in the understanding of the
dynamic properties.

A certain class of disordered mean-field spin-glass models, including,
 e.g., the $p$-spin interaction models, provides a proper description
 for the main features of structural glasses.  These are models
 displaying a stable one-step Replica Symmetry Breaking (1RSB) frozen
 phase \cite{Derrida,Gardner,KWPRA87,CS} very well representing the
 glassy phase of amorphous systems. 
 In that  case, when a
 glass-forming liquid is cooled down below a certain (glass)
 transition temperature, it loses the ability of visiting different
 states (at least on the observation time-scale considered) and this
 is reflected by the loss of entropy.  All the other possible states,
 not selected at the moment of the transition of the liquid to a
 glass, are, anyway, still there from a statistical point of view and
 could, in principle, still be reached on different quenches from high
 temperature or on much larger time-scales. The complexity counts the
 many equivalent states that could have been chosen at the moment of
 the quench.  Depending on the time scale of observation, the
 complexity is thus a purely dynamic concept.

In the mean-field approximation barriers between different metastable
states of finite life-time become infinite in the thermodynamic limit,
eventually breaking the ergodicity.  Complexity can, therefore, be
considered and computed even at the static level.  The presence of
many metastable states can be detected looking at the relaxation,
displaying a dynamical transition with diverging relaxation
time-scales at a given temperature, the {\em dynamical temperature}
$T_{\rm d}$.  Below such transition yet another one can take place, at
the {\em static temperature} $T_{\rm s}$, this time a proper
thermodynamic phase transition to a spin-glass phase.\cite{Gardner}
The complexity displayed by systems whose statics is properly
represented by a 1RSB solution turns out to be extensive.\cite{CS,CGP,CLRpSP}

For what concerns the Sherrington-Kirkpatrick (SK) model,
\cite{SKPRL75} though, that is the mean-field prototype of amorphous
magnets (i.e. actual spin glasses), no dynamical transition precedes
the static one and the frozen phase is a Full Replica Symmetry
Breaking (FRSB) spin-glass phase.  Even though the dynamic behavior
of systems whose static phase is described by a FRSB  or a  1RSB phase
seems to be quite different\cite{Sompolinsky}
the existence of an extensive complexity for the SK model has been put
forward since the very early stages of theoretical modeling of such
materials.

In the last twenty years basically two different approaches have been
presented for the  behavior of the complexity in the SK
model.  The first one was originally introduced by Bray and Moore
\cite{BMan} (BM); there the complexity was analyzed both in the {\em
annealed} approximation, i.e. as the logarithm of the disordered
average of the number of states, and as {\em quenched} average, i.e. the
average of the logarithm.  
The second one was initially proposed by Parisi and Potters \cite{PP}
who showed that the complexity could be obtained by calculating the
partition function of $m$ distinct real replicas of the system
\cite{MPRL95} and provided the connection with the previous BM
formalism by means of a generalization of the {\em two-group
Ansatz}. \cite{BM2g} In that framework they computed an annealed
solution using the so-called 'unbroken' two-group Ansatz.  As we will
see, such an Ansatz turns out to be equivalent to impose a certain
{\em supersymmetry } on the complexity. Indeed, in
Refs. [\onlinecite{MP,CGPM}], explicit computation has led to a
supersymmetric complexity different from the BM one.

Over the years it has become more and more evident that an important
role in the study of the complexity of disordered systems is played by
the so-called Becchi-Rouet-Stora-Tyutin  supersymmetry (BRST-SUSY), first
discovered in the framework of the quantization of gauge theories.
\cite{BRST}
 In the context of stochastic field equations it can be shown that the
 generating functional of correlation functions averaged over
disorder leads to an action displaying BRST SUSY.\cite{ZZ} This
 formally coincides with the average
over the quenched random couplings of the number of solutions of the
mean-field equations for the SK model.
In such a model, the property of BRST-SUSY has been analyzed in
Refs.  [\onlinecite{CGG,CGPM}] and a possible explanation of their
physical meaning has been presented by the authors in Ref.
[\onlinecite{noian}].

In particular, in Ref. [\onlinecite{CGPM}] evidence for a BRST-SUSY
complexity functional was brought about. The difference with the BM
complexity turns out, actually, not to be at the functional level, but simply
the new solution is a second saddle point of the same functional.\cite{noian}
This means that there is just one BRST-SUSY functional (the BM complexity)
displaying (at least) two saddle points, 
one satisfying the BRST-SUSY\cite{CGPM} and the other one not\cite{BMan}.
 Both solutions identify an extensive complexity, computing the
number of solutions of the Thouless-Anderson-Palmer (TAP)
 equations \cite{TAP} in the annealed
approximation.  In Ref. [\onlinecite{noian}] the authors analyzed the two
solutions and their problems reaching the conclusion that both of them
had to be discarded as candidate quenched complexities of the SK model.  So far
for the annealed level.

In the present paper we go further analyzing the quenched complexity
of the SK model in the FRSB scheme, modified
in order to generalize the approach of
Monasson \cite{MPRL95} (applicable to systems with  1RSB
stable static phase).  
The aim  is to see whether or not it
is possible to build a stable (or marginally stable) solution with an
extensive complexity, such as in 1RSB-stable systems like $p$-spin
models.\cite{CS,CGP98,CLRpSP,CLRpIS}

The program we apply amounts to compute the quenched complexity
in the FRSB scheme with proper assumptions.  
Our computation scheme is formally a  particular
case of the one starting from the three-dimensional antiparabolic
equation  proposed by Bray and Moore in Ref. [\onlinecite{BMque}]. The
solution is, however, more
generic than the one obtained, under important constraints,
 by Bray, Moore and
Young. \cite{BMY}
Indeed, our assumptions are less strict then those of 
Ref. [\onlinecite{BMY}], thus allowing for the construction of a
complexity as a function of the free energy.  
Both that solution and the present one are BRST-SUSY
invariant.
The main results of the present paper were anticipated in Ref.
 \onlinecite{CLPRPRL}.

In Secs. \ref{s:rep}-\ref{s:var} we recall the replica computation for the
SK model making the Ansatz of infinite replica symmetry breaking
in the variational form of Sommers and Dupont. \cite{SDJPC84}  
The SK model is stable only in the limit of infinite breaking of the
replica symmetry.\cite{DDGGO} Even then the stability with respect to
overlap fluctuations is only marginal.
In Sec. \ref{s:quenched} we formally generalize the Legendre transform
approach\cite{MPRL95} 
to the case of FRSB stable systems and we modify the replica
computation scheme accordingly, adopting a modified FRSB
scheme in the spirit of Monasson investigation of hidden
states for glass-like models. \cite{MPRL95}
 In Sec. \ref{s:BRSTsol} we show the results for the complexity and
the thermodynamic observables  obtained both by direct
numerical solution and by analytical
expansion around the critical point up to very high order, within 
the modified FRSB scheme.

The stability of the solution is discussed in Sec.  \ref{s:marginal},
where we show that, in the SK model, for any value of
the complexity different from zero, even marginal stability is always
violated.  This means that, requiring BRST-SUSY,
 no exponential growth of the
number of states with the size of the system at any value of the free
energy can be obtained.

What about a BRST-SUSY-{\em{breaking}} FRSB quenched complexity?
Since {\em any} supersymmetric solution would suffer of the same
instability (see Sec. \ref{s:marginal}), in order to give account for
an extensive complexity this remains the only alternative.  In
Sec. \ref{s:nonBRST} we give the most general expressions of the FRSB
complexity and of the differential FRSB equations, without imposing
any symmetry.  Such equations are, basically, those already presented
in Refs. [\onlinecite{BMque,BMY}]. We rederive everything using the
two-group Ansatz formalism, where the breaking of the BRST-SUSY is
more easily carried out and the BRST-SUSY invariant complexity is
described by the unbroken two-group.\cite{BM2g,PP}
 From a purely formal point of view, the equations are equal to those
treated in the study of temperature chaos by means of coupled real
replicas, \cite{CR2} with different boundary conditions.  
Their resolution is currently under investigation.

%%%%%%%%%%%%%%%%%%%%%%%%%%%%%%%%%%%%%%%%%%%%%%%%%%%%%%%

\section{The FRSB SK Complexity}
\label{s:rep}

The Hamiltonian of the SK  model is\cite{SKPRL75}
\BEQ
{\cal {H}}=-\sum^{1,N}_{i<j} J_{ij}\sigma_i\sigma_j
\EEQ
where $N$ is the number of Ising spins $\sigma_i\pm 1$ and $J_{ij}$
are quenched Gaussian random  variables with variance
$\overline{J^2_{ij}}=1/N$ and  mean $\overline{J_{ij}}=0$.
The over-bar denotes the average over disorder.

 Using the replica trick the average free energy  can be expressed in terms 
of the spin-overlap replica matrix  ($a,b=1,\ldots, n$; $n$ being 
the number of replicas):
\BEA
&&\beta f^{\rm rep}=-\frac{\beta^4}{4}
+\frac{\beta^2}{4}\lim_{n\to 0}\frac{1}{n}\sum_{a\neq b}^{1,n} q_{ab}^2 
\label{a_fq}
\\
&&- \lim_{n\to 0}\frac{1}{n}~\log ~\Tr_{\s}\exp\left(\frac{\beta^2}{2}
\sum_{a\neq b}^{1,n} q_{ab}~ \s^a~\s^b\right)
\nn
\EEA
where $\sigma^a$ is the spin variable relative to the replica $a$ and the
trace is taken in the replica space.

In the limit of FRSB the matrix $q_{ab}$ is parametrized by a continuous
non decreasing  function $q(x)$,
$0\leq x\leq 1$, and 
 the free energy functional
(\ref{a_fq}) becomes
\BEA
&&\beta f^{\rm rep}(\beta)= -\frac{\beta^2}{4}\left[
1- 2 q(1) + \int_0^1 dx ~q^2(x)
\right]
\label{f1}
\\
\nn
&&\hspace*{1 cm}-\beta\int_{-\infty}^{+\infty} \d y~P_0(y)~\phi(0,y)
\EEA
where 
\BEQ
P_0(y)=\frac{1}{\sqrt{2 \pi q(0)}}
\exp\left(-\frac{y^2}{2q(0)}\right)
\label{def:P0}
\EEQ
and
$\phi(0,y)$ is the solution evaluated at $x=0$
of the  Parisi antiparabolic differential 
equation:\cite{P80f}
\begin{equation}
\dot\phi(x,y)=-\frac{\dot{q}(x)}{2}\, \left[\phi''(x,y)
+x~\beta\,\phi'(x,y)^2\right]
\label{eqPhi}
\end{equation}
with the boundary condition
\begin{equation}
\phi(1,y)=T\log\left(2\cosh \beta y\right) \ .
\label{Phi1}
\end{equation}
We have used the standard notation, denoting derivatives with respect to
$x$ by a dot and derivatives with respect to $y$ by a prime.

%%%%%%%%%%%%%%%%%%%%%%%%%%%%%%%%%%%%%%%%%%%%%%%%%%%%%%%
\section{Variational Method }
\label{s:var}
             
Functional (\ref{f1}) has to be extremized with respect to $q(x)$.
In order to implement the extremization one can include
the Parisi equation (\ref{eqPhi}) and the boundary condition
at $x=1$, Eq. (\ref{Phi1}) into
 the free energy functional via 
Lagrange multipliers $P(x,y)$: \cite{SDJPC84}
\begin{eqnarray}
&\beta& f^{\rm rep}_{\rm var}(\beta)= \beta f^{\rm rep}(\beta)
  +\beta \int_{-\infty}^{+\infty}\hspace*{-.5 cm} dy\  P(1,y)\,\bigl[\phi(1,y)
\label{eqfrev}
\\
\nn
&& -T \log\left(2\cosh \beta y\right)\bigr]
-\beta\int_0^1\hspace*{-.2 cm}dx\int_{-\infty}^{+\infty}
\hspace*{-.3 cm} dy\ P(x,y) \Bigl\{
\dot\phi(x,y)
\\
\nn
&&\hspace*{2 cm} +
\frac{\dot{q}(x)}{2}\left[
\phi''(x,y) +x~\beta~\left(\phi'(x,y)\right)^2
\right]\Bigr\}.
\end{eqnarray}
The functional
$f^{\rm rep}_{\rm var}$ is stationary with respect to  variations of
$P(x,y)$, $P(1,y)$, $\phi(x,y)$, $\phi(0,y)$ and of the overlap function
$q(x)$.
Variations with respect to $P(x,y)$ and $P(1,y)$  give back
 equation (\ref{eqPhi})
and its boundary condition  (\ref{Phi1}).
Stationarity with respect to variations of $\phi(x,y)$ and $\phi(0,y)$
leads to a partial differential equation for $P(x,y)$ and
to its boundary condition at $x=0$:
\BEA
&&\dot P(x,y) =\frac{\dot q(x)}{2}\, P''(x,y)
-x~\beta~     \dot{q}(x)\, \bigl[P(x,y)\,\phi'(x,y)\bigr]' \ ,
\nn
\\
\label{eqP}
\\
&&P(0,y)=P_0(y)
\label{P0}
\EEA
Eventually, variation with respect to
  $q(x)$ leads to
\BEQ
q(x)=\int_{-\infty}^{\infty} dy\, P(x,y)\, \phi'(x,y)^2 \ .
\label{q}
\EEQ

The Lagrange multiplier $P(x,y)$ can be interpreted as a
 probability distribution.\cite{SDJPC84} The value $q(x)$
of the overlap function can be associated with the  time scale
$\tau_x$ such that for times of order  $\tau_x$ states with an overlap
equal to $q(x)$ or greater can be reached by the system.
In this picture, $P(x,y)$ becomes the probability distribution
of frozen local fields $y$ at the time scale labeled by $x$.\cite{SDJPC84}
 
Starting from Eq. (\ref{q}) and from the interpretation of $P(x,y)$
as distribution of local fields we notice that the function
$m(x,y) = \phi'(x,y)$ represents
the local magnetization over the time-scale $x$. It obeys the equation
\begin{equation}
 \dot m(x,y)=-\frac{\dot q(x)}{2}\, m''(x,y)
 - \beta~x~\dot q(x)\, m(x,y)\ m'(x,y) \ ,
\label{eqm}
\end{equation}
with initial condition
\begin{equation}
m(1,y)=\tanh(\beta y).
\label{eqm1}
\end{equation}

In terms of the local magnetization one can express both the 
equilibrium magnetization and the field-cooled susceptibility of the system:
\BEA
&&m_{\rm eq} = \int_{-\infty}^\infty dy~P(0,y)~m(0,y)
\\
&&
\chi_{FC} = \int_{-\infty}^\infty dy~P(0,y)~m'(0,y)
\EEA

%%%%%%%%%%%%%%%%%%%%%%%%%%%%%%%%%%%%%%%%%%%%%%%%%%%%

\subsection{Marginal Stability of the FRSB Solution}
\label{ss:var1}
In the SK model, with FRSB stable low temperature phase,
 the stability  condition 
can be written as
\BEA
&&1 - 2 q(1)  + \int_{-\infty}^\infty dy~ P(1,y)~ m^4(1,y)
\label{f:stab}
\\
\nn
&&\hspace*{0 cm}=T^2 \int_{-\infty}^\infty dy~ P(1,y)~ [m'(1,y)]^2
leq T^2
\EEA

Deriving equation (\ref{q}) with respect to $x$ we obtain the
  fundamental relation:
\cite{SDJPC84,Som83}

\begin{equation}
\int_{-\infty}^{\infty}dy\ P(x,y) \ [m'(x,y)]^2=1.
\label{rel0}
\end{equation}
and
inserting Eq. (\ref{rel0}) in Eq. (\ref{f:stab}) the latter
turns out to be {\em marginally} satisfied.

Notice  that Eq. (\ref{rel0}) can be used as well to simplify
the expression of some observables. If the external magnetic 
field is zero, the minimum value of $q$, $q(0)$, is zero [$m(0,0)=0$].
That causes Eq. (\ref{P0}) to become $P(0,y)=\delta(y)$.
As a consequence, Eq. (\ref{rel0}) yields
$m'(0,0)^2=1$, 
$m'(0,0)=\pm 1$,
thus implying $m_{\rm eq}=0$ and $\chi_{FC}=1$ in the frozen spin-glass
phase.

%%%%%%%%%%%%%%%%%%%%%%%%%%%%%%%%%%%%%%%%%%%%%%%%%%%%%%%
\section{The Quenched Computation}
\label{s:quenched}

\subsection{Varying the breaking point of the order parameter $q(x)$} 
\label{ss:var2}
The solution of the self-consistency equation (\ref{q}) of the FRSB
solution in the SK model leads to a monotonic continuous order
parameter function $q(x)$,\cite{PJPA80} with $x\in[0,1]$ representing
a continuous ``replica symmetry breaking index'' or, in the dynamical
interpretation of the replica computation,\cite{Sompolinsky,SZPRL81} a
``time-scale label''.  In the absence of external magnetic field
$q(x)$ grows from zero at $x=0$ to its maximum value for $x=x_c$. From
there on it remains constant: $q(x)=q(x_c)=q(1)$, $\forall
x\in[x_c,1]$. The Edwards-Anderson parameter $q_{EA}=q(1)$,\cite{EA}
also called self-overlap, is a physical quantity representing the
overlap of two configurations belonging to the same spin-glass
state.\cite{MPSTV84}
This scenario  naturally emerges solving the Parisi equation
(\ref{eqPhi}) with boundary condition (\ref{Phi1}) imposed at $x=1$.
 The value of $x_c$
is a decreasing function of the temperature, as shown in e.g.
Ref. [\onlinecite{CR}].

\begin{figure}[b!]
\begin{center}
\includegraphics*[width=0.49\textwidth]{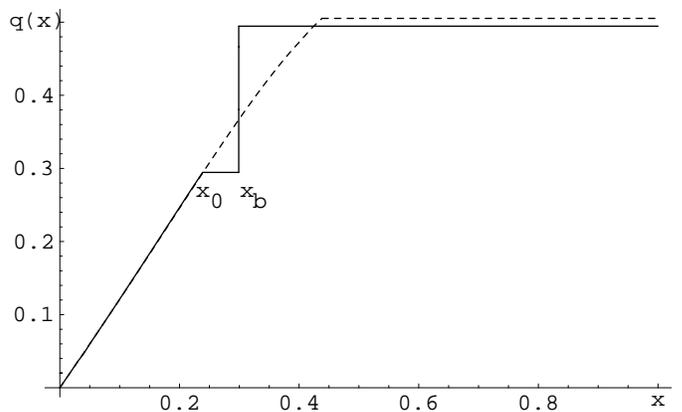}
 \caption{Overlap functions $q(x)$ at $T=0.6$ evaluated resumming the
series expansion in $T_s-T$ ($T_s=1$). 
The dashed line represents the usual FRSB $q(x)$
whereas the continuous line is $q(x)$ computed with an imposed
breaking point $x_b=0.3<x_c=0.530229(1)$. In the latter case $q(x)$
reaches a plateau at a given $x_0$.  Both functions verify the
relationship $\int_0^1q(x)dx=1-T$.}
\label{fig:qPade}
\end{center}
\end{figure}

If, instead than at $x=1$, we set the boundary condition of the Parisi
 equation at a certain $x_b<x_c$, the overlap function extremizing the
 free energy functional develops a discontinuity $\Delta q\equiv
 q(1|x_b)-q(x_b^-)$ at $x_b$, where $q(x_b^-)$ is the value of the
 overlap at the left side of $x_b$ while $q(1|x_b)$ is the value at the
 right side (see e.g. Fig. \ref{fig:qPade}).  In general, the value of
 $q(1|x_b)$ will be different from (larger than) $q_{EA}$.
If the boundary condition (\ref{Phi1}) or, equivalently
(\ref{eqm1}), is imposed at any $x_b$ less than one, but larger than
$x_c$, no change in behavior occurs: the value $q_{EA}$ of the
Edwards-Anderson self-overlap  is reached anyway and the
point at which this happens does not shift. 
  The above construction is a step forward with respect to what was done
in Refs. [\onlinecite{BMY,PP}] where the formal construction was
performed keeping such a difference but the solutions were computed
setting it to zero.

Such a generalization can be directly implemented in the FRSB scheme
 by imposing condition (\ref{b_phi1}) at $x=x_b$ and, at the same time
 by allowing for a discontinuity $\Delta q$. This leads to the
 following free energy functional:

 \BEA 
&\beta f^{\rm rep}_{\rm var}& = -\frac{\beta^2}{4}\left[1-2q(1|x_b)\right]
-\frac{\beta^2}{4}\int_0^{1}dx~q(x)^2 
\\
\nn
&&
-\beta\int_{-\infty}^\infty dy~P_0(y)~\phi(0,y)
\\
\nn
&&+\beta\int_{-\infty}^\infty
dy~P(x_b,y)\left[\phi(x_b,y)-\phi_1(x_b,y)\right] 
\\ \nn 
&&-\beta
\int_0^{x_b}dx~\int_{-\infty}^\infty dy~P(x,y)
\Bigl\{\dot\phi(x,y)
\\
\nn
&&\hspace*{1.5 cm}+\frac{\dot q}{2}\left[
\phi''(x,y) +\beta ~x~\left(\phi'(x,y)\right)^2 
\right]\Bigr\} 
\label{f:Phi}
\EEA 

with the following definitions

\BEA
&&\phi_1(x_b,y)\equiv\frac{1}{\beta~x_b}\log \int {\cal{D}}z
~p(x_b,y,z)
\label{b_phi1}
\\
&&{\cal{D}}z\equiv dz~e^{-z^2/2}
\\
&&p(x_b,y,z)\equiv \left[2\cosh\left(\beta y +\beta z\sqrt{\Delta q}
\right)\right]^{x_b}
\label{f:p1}
\\
&&\Delta q\equiv q(1|x_b)-q(x_b^-)
%\\
%&&P_0(y)\equiv \frac{1}{\sqrt{2\pi q(0)}}\exp\left(-\frac{y^2}{2~q(0)}
%\right)
\EEA

If $x_b >x_c$, $\Delta q$ is zero  and  
 the boundary condition (\ref{b_phi1}) becomes
\BEQ
\nn
\phi_1(x_b,y)=\frac{1}{\beta}\log 2 \cosh \beta y=\phi(1,y) \
 \hspace*{.1 cm}; \ \  \forall~ x_b\ge x_c
\EEQ
This way we recover the usual unconstrained FRSB solution of the SK model.

It is worth noticing that solving Eq. (\ref{eqPhi}) simply
with boundary condition
(\ref{Phi1}) imposed at at $x_b<x_c$ just leads to same result, 
i.e. to the same
discontinuous $q(x)$. 
The choice we make here of introducing $\Delta q$ already in the boundary 
condition  just makes practical computation easier.

\begin{figure}[t!]
\begin{center}
{
\includegraphics*[width = 0.49\textwidth]{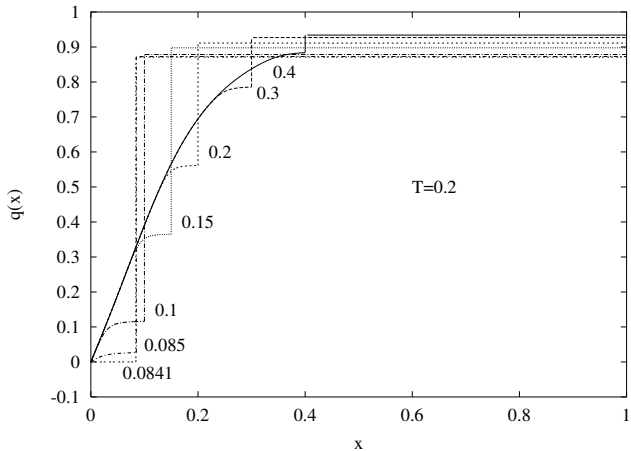}
}
\caption{Order parameter of the FRSB solution obtained by numerically
solving the Eqs. (\ref{eqP}),(\ref{eqm}), (\ref{q}),(\ref{f:q1}), with
modified boundary condition for the Parisi equation, imposed at
$x=x_b$ instead than at $x=1$. $T=0.2$. For $x_b<x_c=0.54173(1)$ a
discontinuity occurs.  Notice that for $x_b= 0.0841$ ($x_{\rm 1RSB} =
0.0845(2)$) the continuous part of $q(x)$ goes to zero in the whole
interval $[0,x_{\rm 1RSB}]$.  At $x_b=x_{\rm 1RSB}$ the free energy of
the 1RSB solution becomes lower then the relative one in  the FRSB
solution. All overlap functions satisfy $1-\int dx~q(x) = T$.  }
\label{fig:qNum}
\end{center}
\end{figure}

For numerical purposes the study of the equation (\ref{eqm}) for the
local magnetization $m(x,y)\equiv \phi'(x,y)$ is better suited.  For
$x\leq x_b$, the stationarity Eqs. (\ref{eqm}), (\ref{eqm1}) are
replaced by

\BEQ
m(x_b,y)=m_1(x_b,y)= \left<\tanh\left(\beta y +
\beta z \sqrt{\Delta q}\right)\right>
\label{f:m1xc}
\EEQ
where the average is defined as
\BEQ
 \left< o(y,z) \right> \equiv \frac{\int {\cal{D}}z~ o(y,z)~ p(x_b,y,z)}
{\int {\cal{D}}z ~ p(x_b,y,z)}
\label{f:ave}
\EEQ
and  the self-consistency equation (\ref{q})
for $q(x)$ is replaced by
\BEA
&&q(x)=\int_{-\infty}^\infty \hspace*{-.3 cm}dy~P(x_b,y)\left<
\tanh^2\left(\beta y+\beta z\sqrt{\Delta q}\right)\right>
\\
\nn
&&\hspace*{4.2 cm}=q(1|x_b) \ 
\ \ \ \mbox{if $x>x_b$}
\label{f:q1}
\\
&&q(x)=\int_{-\infty}^\infty \hspace*{-.3 cm}dy~P(x,y)\left[m(x,y)\right]^2
\hspace*{1cm} \mbox{if $x<x_b$}
\EEA
We notice that Eq. (\ref{f:m1xc}) for $x_b>x_c$, i.e. $\Delta q = 0$,
becomes Eq. (\ref{eqm1}).

  In Fig. \ref{fig:qPade} we plot $q(x)$ at $T=0.6$ and $x_b<x_c$ by
means of series expansions around the critical point and subsequent
Pad\`e resummation.  In Fig. \ref{fig:qNum} we show $q(x)$ at $T=0.2$
for different values of $x_b<x_c$, obtained by numerical resolution of
the Parisi equation.  Notice that for $x_b$ less than a particular
value (we can call it $x_{\rm 1RSB}$) $q(x)$ becomes a step function.
This happens because at $x_b\leq x_{\rm 1RSB}$ the free energy of the 1RSB
solution becomes lower then the relative one  for the FRSB
solution. All overlap functions do satisfy the relation $1-\int
dx~q(x) = T$.
%%%%%%%%%%%%%%%%%%%%%%%%%%%%%%%%%%%%%%%%%%%%%%%%%%%%%%%

\subsection{The Coupled Real Replicas Approach\cite{MPRL95,PP} Generalized
to FRSB-stable Spin-Glasses}
\label{ss:var3}

The equations discussed in the previous section can be obtained by
considering $m$ equivalent copies (real replicas) of the system
interacting through an infinitesimal coupling.

If $Z_{J}$ is the partition function for a given realization of the
couplings $J_{ij}$, the starting point for such a computation is the
replicated partition function $\overline{Z_{J}^{\ n m}}$, where $n$ is
the number of replicas of a single copy of the system and $m$ are the
real replicas of the system interacting between each other through an
infinitesimal coupling. Afterward we send $n$ to zero, keeping $m$
finite\cite{MPRL95,PP} (see also Sec.  \ref{s:nonBRST}).

Because of the interaction between the $m$ copies the resulting
solution in the space of $n\times m$ replicas is not just the one we
would get for $n$ replicas repeated $m$ times
%, that is, the one we
%would obtain computing $\left(\overline{Z_{J}^{n}}\right)^m$.  
In other words, the values of the elements of the overlap-matrix of
each copy are not the same one would get in absence of the coupling.

Using the permutation symmetry of the $m$ replicas it is possible to
perform proper permutations of lines and rows of $q_{ab}$ in order to
get the highest overlap values grouped together in a diagonal block of
dimension $m$.  Sending $n \to 0$  leads to a finite $m\in [0,1]$,
 representing a sort of density of the number of copies.

Setting a value $x_b<x_c$ at which we impose the boundary condition of
 the FRSB equations, is equivalent to cut the refinement of replica symmetry
 breaking inside the diagonal blocks of the Parisi matrix at $q(x_b)$.
 Beyond this point no further breaking occurs and all the elements of
 the matrix inside the diagonal blocks take the same value
 $q(1|x_b)>q(x_b)$. Such a quantity is always something less than the
 Edwards-Anderson self-overlap [$q(1|x_b)<q(1|x_c)$]. This cane be easily 
understood since it is a
 sort of average of the monotonously increasing overlap function over
 the $q$ values labeled by $x>x_b$.

Identifying $x_b$ with $m$,
when the 'copies density' $m=x_b$ is chosen higher than the critical
value $x_c$ 
%at which the order parameter of the model solved with the
%unconstrained scheme continuously reaches the maximum value
%$q(1)=q_{EA}$, 
the refinement is complete and the solution coincides
with the one without discontinuity.
%Technically speaking, imposing the boundary condition 
%$m(1,y)$ at $x=1$ [Eq. (\ref{eqm1})]
%or at any other $x\in[x_c,1[$ yields precisely the 
%same result. The largest overlap is  $q_{EA}$ in any case
%and the shapes of the monotonous continuous part of $q(x)$ below $x_c$
%do coincide exactly.

%%%%%%%%%%%%%%%%%%%%%%%%%%%%%%%%%%%%%%%%%%%%%%%%%%%%%%%
\subsection{Complexity as Legendre Transform of the FRSB 
Replica Thermodynamic Potential}
\label{ss:var4}

Making the identification $m=x_b$ and looking at the analogy with systems 
displaying a stable 1RSB frozen phase we can generalize the approach 
or Ref. [\onlinecite{MPRL95}] to FRSB systems, defining  also in this
  case the complexity $\Sigma$ as the Legendre Transform of  
replicated  free energy
 $f^{\rm rep}$ as
% using the breaking point $x_b$,
\BEQ
\Sigma(f)=\max_{x_b}\left[\beta x_b f- \beta x_b f^{\rm rep}\right]
\label{f:SLeg}
\EEQ
with the  free energy $f$ and $\beta x_b$ 
as Legendre conjugated variables:
\BEA
&&\beta x_b=\frac{\p~\Sigma}{\p f}
\label{f:bx1}
\\
&&f=\frac{\p~ x_b f^{\rm rep}}{\p x_b}
\label{f:f}
\EEA
Exploiting Eq. (\ref{f:f}), 
the complexity can be equivalently expressed as

\BEQ
\Sigma(f)=\left.\beta~x_b^2 \frac{\p f^{\rm rep}}{\p x_b}\right|_{x_b(f)}
\label{f:sigma}
\EEQ
where the relation $x_b(f)$, or vice-versa $f(x_b)$, is yielded,  e.g., by
Eq. (\ref{f:f}). 
If
 $x_b< x_c$,
%  the non zero $\Delta q$ displayed by
% the overlap function makes the stationarity of 
$f^{\rm rep}$ is not stationary with respect to the breaking parameter
$x_b$ and, therefore, $\Sigma(f) \neq 0$ in some $f$-interval $[f_0,
f_{\rm th}]$.  As $x_b\to x_c$ the unconstrained FRSB solution is
recovered, thus $f(x_c)=f_{\rm eq}$ and it also coincides with the
lower band-edge $f_0$ at which the complexity goes to zero.

%%%%%%%%%%%%%%%%%%%%%%%%%%%%%%%%%%%%%%%%%%%%%%%%%%%%%%%

\section{The FRSB BRST-SUSY Solution and its Properties.}
\label{s:BRSTsol}

The explicit expressions for the   free energy  and the complexity 
as functions of $x_b$ are obtained from 
Eqs.  (\ref{f:f}) and (\ref{f:sigma}) and respectively: 
\BEA
&&
f(x_b)=
\frac{\beta x_b}{4}\left[q^2(1|x_b)-q^2(x_b^-)\right]
\\
\nn
&&
\hspace*{.0cm}+\int_{-\infty}^\infty dy~P(x_b,y)
	\left[	\phi_1(x_b,y)
		-\frac{1}{\beta x_b}\left<
		\log p(x_b,y,z) \right>
	\right]
\\
\nn
&&\hspace*{1.0cm}-\frac{\beta}{4}\left[1-2q(1|x_b)
+\int_0^{1}dx~q^2(x)\right]-\phi(0,0)
\\
&&\Sigma(x_b)=\frac{(\beta x_b)^2}{4}\left[
q(1|x_b)^2-q(x_b^-)^2\right]
\\
&&
\hspace*{.0cm}
+\int_{-\infty}^\infty dy~P(x_b,y)
\left[\beta x_b~\phi_1(x_b,y)-\left<\log p(x_b,y,z)\right>
\right]
\nn
\EEA

In particular, the  free energy can be written as 

\BEQ f= e - T s
\EEQ
 where $e$ is the internal energy density and $s$ the entropy
contribution inside a single spin-glass state.

Thermodynamic relations are valid for any $x_b$ and, therefore, the 
expressions for energy and entropy are
\BEA
&&u(\beta) = \frac{\p \beta f^{\rm rep}}{\p \beta}=
-\frac{\beta}{2}\left[1-\int_0^1dx~q^2(x)\right]
\\
&&s(\beta,x_b) = \beta\left[\beta \frac{\p f^{\rm rep}}{\p \beta}\Bigr|_{x_b} 
			- x_b\frac{\p f^{\rm rep}}{\p x_b}\Bigr|_\beta\right]=
\\
\nn
&&
-\frac{\beta^2}{4}\left[1-q(1|x_b)\right]^2
-\beta^2 x_b~ \Delta q\left[1-(1-x_b)q(1|x_b)\right]
\\
\nn
&&
+\int_{-\infty}^\infty dy~P(x_b,y)\left[\frac{1}{x_b}\left<
\log p(x_b,y,z)\right>
-y ~m_1(x_b,y)\right]
\EEA
where we used the fact that the replica  entropy 
$\beta^2\p f^{\rm rep}/\p \beta$ is equal to
$s+\Sigma/x_b$.
Notice that, as $\Delta q \to 0$, the expression of the entropy tends to the
usual FRSB one.

The solution we are considering satisfies the BRST SUSY.
\cite{CGG,CGPM} We will show later on (Sec. \ref{s:nonBRST}) how this
can be simply proved, starting from a general formulation involving
more than one order parameter and reducing to the present one when
certain relations among them, stating the supersymmetry, are
imposed. 
%For the time being, in this section we show the behaviour of
%the candidate complexity for the SK model making use of the FRSB with
%a built-in discontinuity $\Delta q$ in the overlap function
%(FRSB$_{\Delta q}$ Ansatz).}

In the following we present our results for the complexity
in the FRSB$_{\rm \Delta q}$ Ansatz obtained both by 
 numerical resolution of   Parisi equation and by
analytical expansion around $T_s$.

%%%%%%%%%%%%%%%%%%%%%%%%%%%%%%%%%%%%%%%%%%%%%%%%%%%%%%%

\subsection{Numerical Resolution of FRSB Equations 
with Pseudo-Spectral Integration}
\label{ss:num}

We performed an accurate numerical study of Eqs. (\ref{eqP}),
(\ref{eqm}) with imposed breaking at $x_b<x_c$,
 exploiting the pseudo-spectral numerical
integration method used, e.g., in Refs.  [\onlinecite{CLP,CR,CL}].  As
a comparison we also performed the study of the 1RSB solution and its
complexity (analyzed and discussed in Ref. [\onlinecite{CGPM}]) for
the same values of the temperature.
 
\begin{figure}[b!]
\begin{center}
{
\includegraphics*[width = 0.49\textwidth]{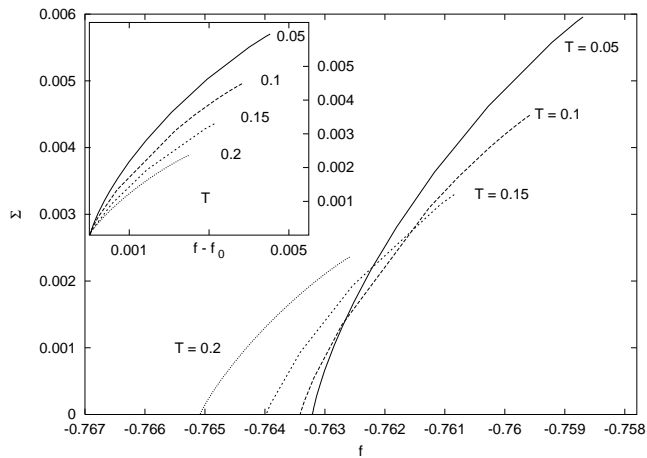}
}
\caption{Complexity vs. $f$ at different temperatures. The interval
$[f_0,f_{\rm th}]$ decreases as $T$ is increased toward the critical
temperature $T_s=1$.}
\label{fig:S_T_mul}
\end{center}
\end{figure}

In figure \ref{fig:S_T_mul} we show the behavior of the FRSB$_{\Delta
q}$ complexity as a function of the free energy.   We  observe that in
the SK model the  complexity (\ref{f:sigma}) is non-zero in a given  interval
$f\in[f_0,f_{\rm th}]$
% even in absence of the dynamic transition
%taking place  in 1RSB stable systems (see
%e.g. Refs. [\onlinecite{CS,CGP98,CLRpIS}]).
As shown in the inset of  Fig. \ref{fig:S_T_mul}, the interval of
free energies $f$ in which the complexity is non zero
 increases as $T$ decreases. This is also the
dominion of variation of $f^{\rm rep}(f)$.

\begin{figure}[th!]
\begin{center}
\includegraphics*[width=0.49\textwidth]{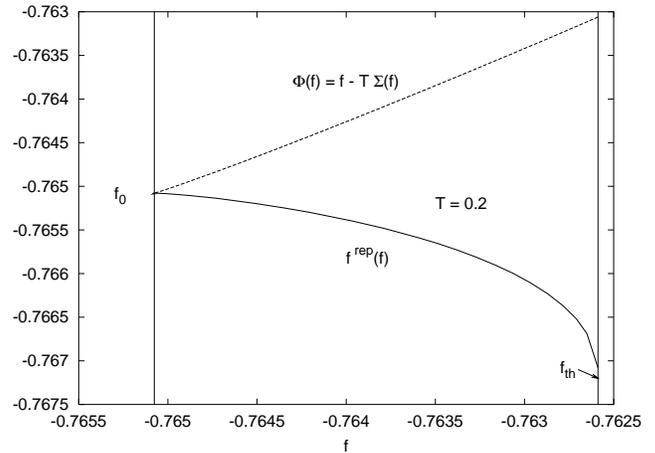}
\caption{Behavior of $f^{\rm rep}(f)$ and $\Phi(f)= f - T \Sigma(f)$.
The two functions converge to $f_0$ as $f\to f_0$ (the equilibrium
free energy), where the complexity goes to zero.}
\label{fig:f_Phi_f}
\end{center}
\end{figure}

\begin{figure}[th!]
\begin{center}
\includegraphics*[width=0.49\textwidth]{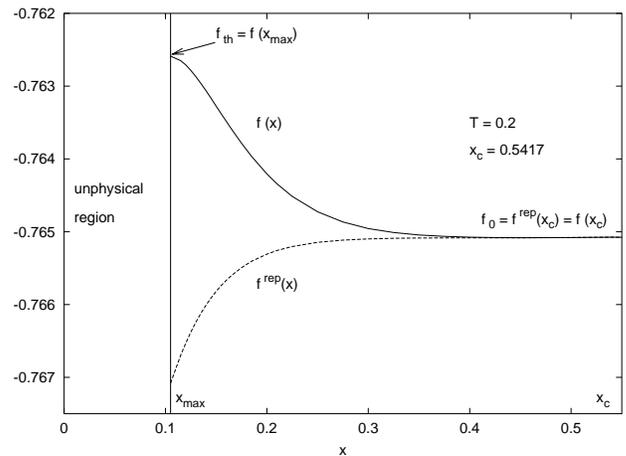}
\caption{Replicated free energy $f^{\rm rep}$ and state free energy
$f$ as functions of $x$ in the region $[x_{\rm max },x_c]$ at
temperature $T=0.2$.  The maximum value of $f^{\rm rep}$ ($f_{\rm eq}$
occurs for $x = x_c$, where it coincides with the lowest value of
$f$.}

\label{fig:Phi_f_x}
\end{center}
\end{figure}

Above the static transition between paramagnet and FRSB spin-glass, at
temperature $T_s$, the complexity is zero. Below $T_s$, if we look at
the behavior of the replicated thermodynamic potential $f^{\rm rep}$
both as a function of the state free energy $f$ and as a function of
the breaking parameter $x_b$ (see
Figs. \ref{fig:f_Phi_f} and \ref{fig:Phi_f_x}) we can see that the
extremal value of the replicated free energy is always at the lowest
value of the free energy $f$: $f_0=f(x_c)=f^{\rm rep}(x_c)$.  This is
the value of vanishing complexity and coincides with the equilibrium
thermodynamic free energy $f_{\rm eq}$.

\begin{figure}[th!]
\begin{center}
\includegraphics*[width = 0.49\textwidth]{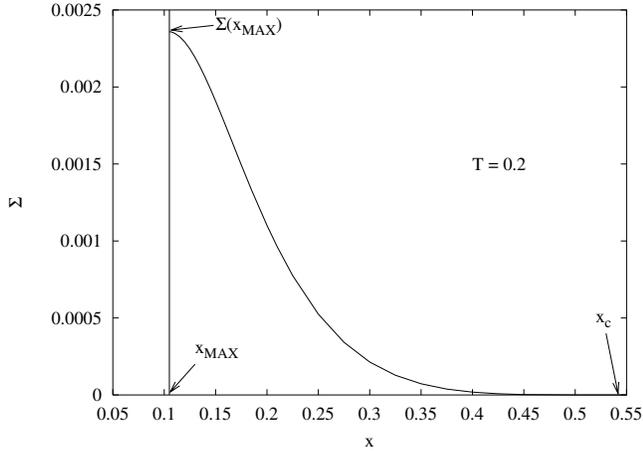}
\caption{ Complexity vs. $x$ at $T=0.2$. $x_{\rm max }=0.105(1),
x_c=0.54173(1)$. }
\label{fig:S_x}
\end{center}
\end{figure}

\begin{figure}[th!]
\begin{center}
\includegraphics*[width = 0.49\textwidth]{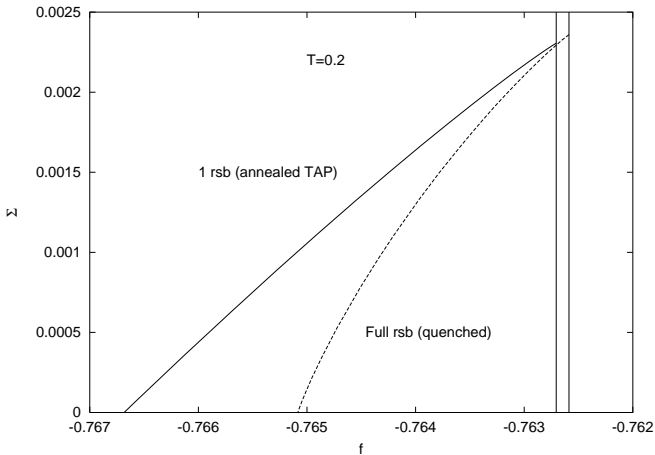}
\caption{FRSB$_{\Delta q}$ and 1RSB complexities versus $f$.}
\label{fig:FRSB_S_f}
\end{center}
\end{figure}

\begin{figure}[th!]
\begin{center}\includegraphics*[width = 0.49\textwidth]{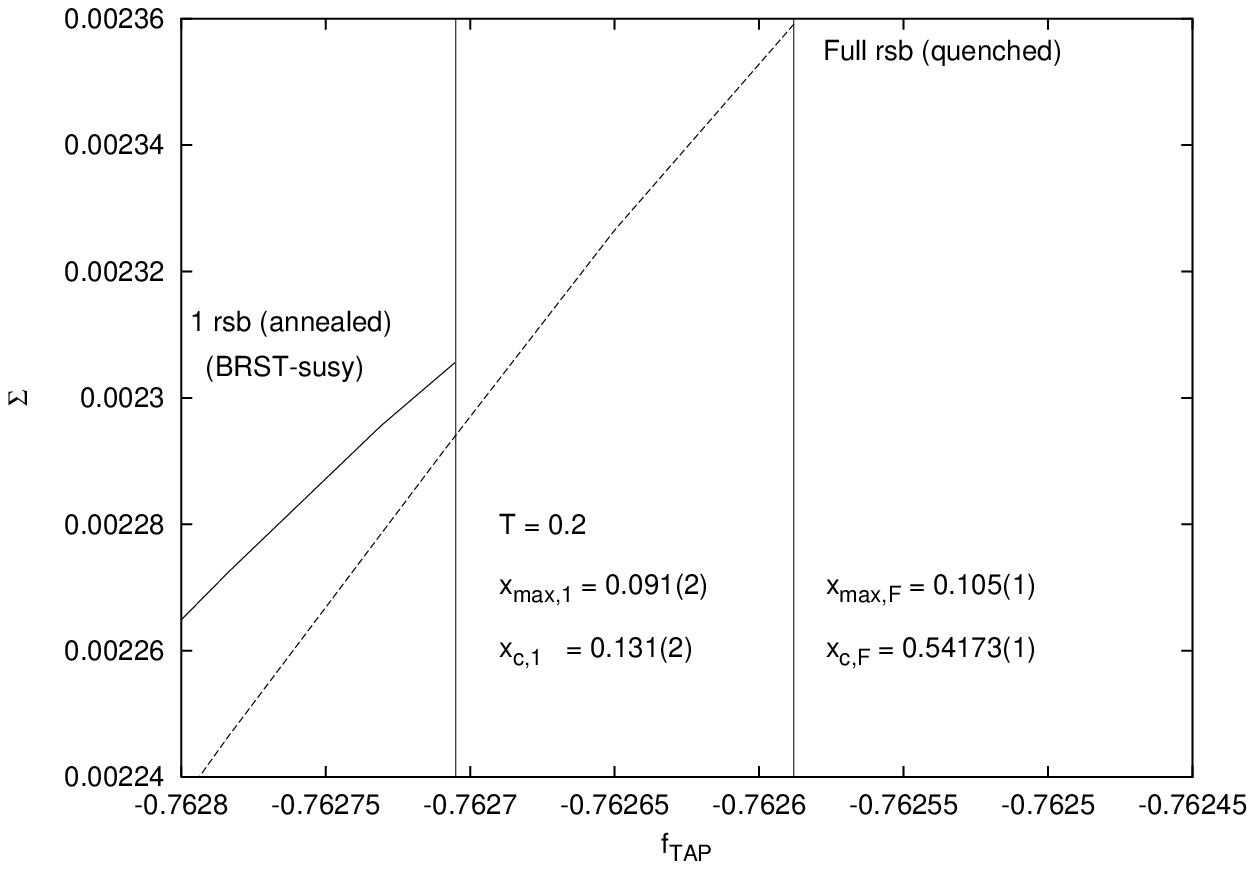}
\caption{FRSB$_{\Delta q}$ and 1RSB complexities. Detail around the 
threshold values of the TAP-state  free energy. }
\label{fig:FRSB_S_f_det}
\end{center}
\end{figure}

We define $x_{\rm max}$ the value of 
$x_b$ corresponding to $f_{\rm th}$ and to the maximum complexity
(see Figs. \ref{fig:Phi_f_x}-\ref{fig:S_x}):
In figure \ref{fig:Phi_f_x} according to the interpretation mutuated
from 1RSB-stable systems, \cite{CGPM} we can see that for any $x_{\rm
max}<x<x_c$ the values of $f$ are increasing as $x_b$
decreases: they should be the free energy values of the most
significant states.  Their number is exponentially large and equal to
$\exp\left[N\Sigma(f)\right]$.  
Beyond $x_{\rm max}$ the solution is said 'unphysical' ($f(x)$
decreases with $x$).

On the contrary, as $x$ decreases from $x_c$ to $x_{\rm max}$, the
total free energy $f^{\rm rep}$ decreases. Indeed $f^{\rm rep}$ is the
free energy computed by the replica trick and, thus, it has to be
maximized with respect to variations of the order parameter values.
Adopting the usual, unconstrained, scheme of computation the maximum
value of the SK free energy, $f^{\rm rep}=f_{\rm eq}$ occurs for $x_b\geq
x_c$.
If we compute it at any other value of $x_b$, $f^{\rm rep}$ is less than
this.  In figure \ref{fig:Phi_f_x} the behaviours of both $f^{\rm
rep}$ and $f$ as functions of the imposed breaking parameter are shown
at $T=0.2$.
% From a dynamical point of view $f_0$ corresponds to the
%free energy value at longest time scale, i.e.  the equilibrium value,
%whereas the $f_{\rm th}$ represent the threshold free energy of the
%most relevant states to which the system evolves.

Having analized in detail the behaviours of both the complexity and
the replicated free energy we summarize that 
(a) the maximum of
$f^{\rm rep}(f)$ or, equivalently, the minimum of
$\Phi(f)=f-T\Sigma(f)$, is given by the extremum $f_0$ of the
$f$-support (the equilibrium free energy in the FRSB stable SK model),
for any temperature, whereas in models where a $T_{\rm d}\neq T_{\rm
s}$ exists (1RSB stable systems) such a stationary point is somewhere
in the middle of the support, 
going to $f_0$ as temperature {\em decreases}
\cite{MPA99, Coluzzi}
(b) the support of the curve $\Sigma(f)$ increases from zero as
temperature is decreased from the critical point (Fig. \ref{fig:S_T_mul}),
 whereas in 1RSB
stable systems, below  $T_s$, the $f$-support
decreases as $T$ is lowered.

Such apparent contradictions can be solved studying the Ising $p$-spin
model, which has both a 1RSB and a FRSB phase.\cite{Gardner} Such an
analysis will be presented elsewhere. \cite{CLRpIS}

\begin{figure}[t!]
\begin{center}
\includegraphics*[width = 0.49\textwidth]{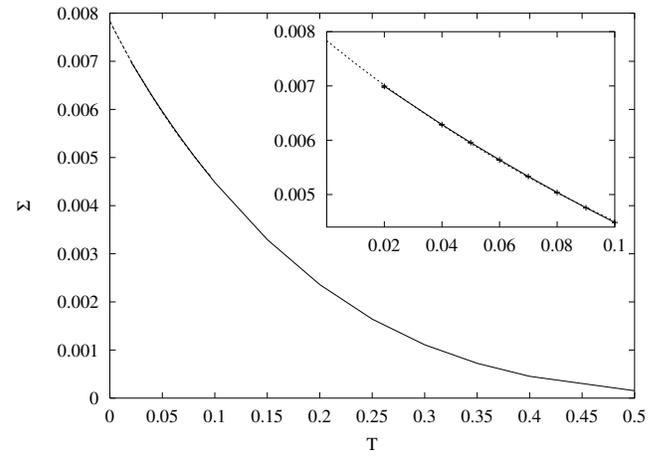}
\caption{ Complexity $\Sigma$ versus T for the FRSB$_{\Delta q}$ solution.
% $\Sigma(0)=0.00778(5)$
The dashed part of the curve is obtained by a fit.}
\label{fig:S_T}
\end{center}
\end{figure}

A closer analysis of the complexity of the SK model at a given
temperature as a function of the free energy rises yet another, more
serious, problem.  In figure \ref{fig:FRSB_S_f} the BRST-SUSY quenched
complexity is plotted as a function of $f$ together with the BRST
annealed complexity obtained with a 1RSB $q(x)$ \cite{CGPM, noian} at
$T=0.2$.
  Convexity implies that the annealed average must be larger than the
quenched one and this must hold at any value of $f$.  As we can see
from the figures \ref{fig:FRSB_S_f}-\ref{fig:FRSB_S_f_det}, the
threshold value $f_{\rm th}$ in the quenched case (vertical line on
the right side in Fig. \ref{fig:FRSB_S_f_det}) is greater than in the
annealed one (left side vertical line).  This leads to a violation of
the convexity because in the interval of $f$ values between the
annealed and the quenched threshold free energies the annealed
complexity is zero while the quenched one is finite.  Also
their maximum values violate the convexity requirement as it is shown
in Fig. \ref{fig:FRSB_S_f_det}. In Fig.  \ref{fig:S_T} the maximum
value of $\Sigma$ is shown versus temperature.  In the limit $T\to 0$
we get, for the maximum value of the complexity, the limit
\BEQ \lim_{T\to 0} \Sigma^{\rm max}(T) = 0.00778(5) \EEQ 
larger than the value obtained in the annealed approximation
for the SK model ($\Sigma^{\rm max}(0)=0.0073$),\cite{CGPM} corresponding to a
1RSB statics in absence of external fields.

%%%%%%%%%%%%%%%%%%%%%%%%%%%%%%%%%%%%%%%%%%%%%%%%%%%%%%%
\subsection{Analytical Resolution with Pad\`e Resummation of Series
Expansion}
\label{ss:pade}
As we have seen (figure \ref{fig:qPade}), $q(x)$ is a continuous
monotonous function for $x<x_b$.  At some $x_0<x_b$ it develops a
plateau such that $q(x)=q(x_0)$ for $x_0 \leq x \leq x_b$, in
particular $q(x_b^-)=q(x_0)$.
Setting $x_b$ near the break point of the Parisi solution, $x_c$, 
and defining their difference as $\epsilon = x_c-x_b$, we derive the
following relationships valid at any temperature:
\begin{eqnarray}
\Delta q & = & O(\epsilon)
\\
x_0 & = & x_c-O(\epsilon)
\\
q(1) & = & q_c(1)-O(\epsilon^2)
\\
q(x_0) & = & q_c(1)-O(\epsilon)
\\
q(x) & = & q_c(x)+O(\epsilon^4) \ \ \ x<x_0
\\
f^{\rm rep}(\epsilon) & = & f^{\rm rep}_c+O(\epsilon^5)
\end{eqnarray}
where the subscript $c$ stays for the case $x\geq x_c$, i.e. the
unconstrained FRSB solution of the SK model. In particular,
$q_c(1)=q_{\rm EA}$ and $f_c^{\rm rep}=f_{\rm eq}$.

From the derivative of Eq. (\ref{f:f}) we see that
\begin{equation}
\left[ 2{\partial f^{\rm rep} \over \partial x_b}+x_b{\partial^2
f^{\rm rep} \over \partial x_b^2} \right]_{x_b=x_{\rm max}} =0
\end{equation} 
the state free energy $f$ takes its maximal value as a function of
 $x_b$; as a consequence, smaller values of $x_b$ have no physical
 meaning.
Notice that at $x_b=x_{{\rm max}}$ the complexity $\Sigma (x_b)$ is
 also maximal and $\Sigma^{\rm max}=\Sigma (x_{\rm max})$ represents
 the total complexity at that temperature.

Decreasing $x_b$, the value $q(x_0)$ of the lowest plateau
decreases; in particular, at
 $x_b=x_{\rm 1RSB}$ we have $\Delta q=q(1|x_b)$, that is the FRSB solution free
 energy becomes higher than the one of the 1RSB function with $q_0=0$
 and $q_1=q(1|x_b)$.  However, $x_{\rm 1RSB}<x_{\rm max}$ and, therefore, it
 is always outside 
 the ``physical''  range of $x_b$'s.

\begin{figure}[t!]
\begin{center}
\includegraphics*[width=0.49\textwidth]{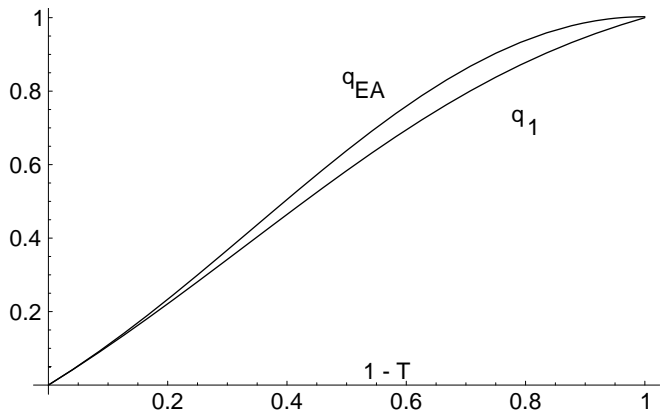}
 \caption{higher curve: $q_{EA}$ as a function of $\tau=1-T$; lower
 curve: self-overlap $q(1|x_{\rm max})$ of the states with maximal
 complexity as a function of $\tau=1-T$, note the finite slope at
 $T=0$.}
\label{figure3}
\end{center}
\end{figure}

We have computed the function $q(x)$ and the various quantities of
 interest in a series expansion in powers of $\epsilon$ and of the
 reduced temperature $\tau=T_s-T$ ($T_s=1$) at very high orders
 through the methods described in Ref. [\onlinecite{CR}].  The various
 series turn out to be asimptotycs but they can be resummed {e.g.}
 through Pad\'e approximants yielding very precise results in the
 whole low temperature phase.  
For the total complexity $\Sigma^{\rm max}(T)$ at temperature $T$ we have the
 following series expansion:

\BEQ
 \Sigma^{\rm max}(T) = {\frac{{{\tau }^6}}{81}} - {\frac{2\,{{\tau
 }^7}}{81}} + {\frac{187\,{{\tau }^8}}{1215}} - {\frac{9938\,{{\tau
 }^9}}{10935}} +O\left(\tau^{10}\right) 
\label{S_tau}
\EEQ  
Higher orders show that this series is strongly asymptotycs 
(see Appendix),
 but Pad\'e approximants of it give
 good results, in particular we have
\begin{equation}
\lim_{T\rightarrow 0}\Sigma^{\rm max}(T)=0.0077(2)
\end{equation}
in agreement with the result obtained by numerical evaluation of the
differential equations of the preceding section.  
If we compare the quenched complexity with the annealed one,
\cite{BMan,CGPM,noian} we can observe that, at any order in $\tau$,
the convexity property is conserved up to the threshold energy value
of the annealed solution, $f^a_{\rm th}$ (compare with figure
\ref{fig:FRSB_S_f_det}), while it is violated soon after.  At the
threshold $f_{\rm th}$ the difference between the quenched and the
annealed complexity is

\BEQ 
\Sigma(f_{\rm th}^a)-\Sigma_a(f_{\rm th}^a) = -\frac{
3 + 2\,{\sqrt{2}}}{2187} \tau ^9
+O\left(\tau^{10}\right)<0 
\EEQ

On the contrary,
the difference between the maximum value of the two complexities is,

\BEQ 
\Sigma^{\rm max}(\tau)-\Sigma_a^{\rm
 max}(\tau)={\frac{{{\tau }^8}}{243}} - {\frac{2\,{{\tau }^9}}{135}} +
 {\frac{3394\,{{\tau }^{10}}}{54675}} +O\left(\tau^{11}\right)
\EEQ 
that is positive to the
 leading order, thus violating convexity.
The above expansions up to very high order is reported in Appendix.

The self-overlap of the states with highest free energy  
is always smaller than the Edwards-Anderson parameter $q_{EA}$, see
figure \ref{figure3}, in particular near $T=0$, the self-overlap of
the states with maximal complexity is
\begin{equation}
q(1|x_{\rm max})=1-0.4 T+O(T^2)
\end{equation} 
while the Edwards-Anderson parameter has a $T^2$ dependence.

%%%%%%%%%%%%%%%%%%%%%%%%%%%%%%%%%%%%%%%%%%%%%%%%%%%%%%%
\section{Violation of Marginal Stability}
\label{s:marginal}

To solve the above mentioned problems the analysis must, then, be
deepened. The behaviour of the complexity in itself not being enough
to understand the properties of the system.  We go, then, further 
analyzing  its stability in the FRSB$_{\Delta q}$
scheme of computation.

\subsection{The Plefka Criterion}

A necessary condition  for the physical relevance of
 the solution is the Plefka criterion,\cite{Ple0,Ple1} stating that the
quantity
\BEQ
x_{\rm P}=1 - \beta^2 {\overline{\left(1-m^2\right)}},
\label{xP}
\EEQ
where the overbar denotes the average over quenched disorder,
must be positive or zero.
Such quantity coincides with the replicon eigenvalue
$\Lambda_R$ in the replica computation.

In order to compute $x_{\rm P}$ we need the quantity $P(x_b^+,y)$. 
To this aim one can  consider $q$ as the
independent variable instead of $x$.  The discontinuities in $q(x)$
become plateaus of $x(q)$ and {\it viceversa} (see Figure \ref{x_q}).
The functions $P(q,y)$ and $m(q,y)$ satisfy the differential equations
\begin{equation}
\label{SP2}
\partial_q m(q,y)=-\frac{1}{2}m''(q,y)
-\beta~x(q)~m(q,y)~m'(q,y)
\end{equation}
\begin{equation}
\label{SP3}
\partial_q P(q,y)=\frac{1}{2}P''(q,y)
-\beta~x(q)[m(q,y)P(q,y)]'
\label{equaP}
\end{equation}
with initial conditions (in the absence of a magnetic field) \begin{eqnarray}
m(q_1,y) & = & \tanh (\beta y)\label{condini} \\
P(0,y) & = & \delta (y)
\end{eqnarray}
where $q_1=q(1|x_b)$.
Notice that the previous equations do not depend on the derivatives
of $x(q)$; therefore we can safely use them  also where $x(q)$ is
discontinuous. This would not be true if we had used $q(x)$ instead of
$x(q)$.  

By solving these equations at
fixed $x=x_b$ we obtain:
\BEA
&&\int dy~P(q_{1},y)~ \tanh^k \beta y =\int dy ~\tanh^k y 
\\
\nn
&&\hspace*{1 cm}\times 
\frac{\int~ dy'~ G_{\Delta q}(y-y')~P(q_b,y')~\cosh ^{x_b} (\beta y)}
{\int dy''~ G_{\Delta q}(y''-y')~ \cosh ^{x_b}(\beta y''-\beta y') }
\EEA
where $G_\Delta(z)=\exp[-z^2/(2\Delta)]$ and $q_b\equiv q(x_b^-)$.
By a simple change of variables the last expression can be rewritten as:
\BEA
&&\int dy~P(q_{1},y) \tanh^k \beta y 
\\
\nn
&&\hspace*{2cm}=\int dy~P(q_b,y) \left< \tanh^k
\beta (z\sqrt{\Delta q}+y) \right>
\EEA
where the average $\left<\left(\ldots\right)\right>$ 
is defined in  Eq. (\ref{f:ave}), so that 
the Plefka parameter finally reads:
\BEA
&&x_{\rm P}= 1-\beta^2\Bigl[1 - 2 q_1
\label{xp}
\\
\nn
&&\hspace*{1.2 cm} +
\int dy~P(q_b,y) \left< \tanh^4\left(
\beta y +\beta z\sqrt{\Delta q}\right) \right> \Bigr]
\EEA

In order to derive the Plefka parameter $x_{\rm P}$ 
 we have to compute disorder averages of the
single-site magnetization $\overline{m^k}$.  They can be expressed in
terms of the cavity field $y$ which is the field induced by $N$ spins
on the $N+1$-th spin $s_0$, averaged in the absence of the spin $s_0$.
\cite{MPV}  Introducing the distribution function $P(y)$ of the
cavity field, averaged with respect to the states and to the disorder,
we have:

\begin{equation}
\overline{m^k}=\int dy~ P(y)\tanh^k \beta y
\end{equation}
It can be shown\cite{ThoThoChoSheSom86}  that the
distribution of the cavity field at equilibrium is simply related to
the function $P(x,y)$  introduced to compute the Parisi solution
$q(x)$.\cite{SDJPC84,S85,CR}

We now briefly discuss
how the same results can be obtained using the cavity method.
According to the Parisi picture the average number of equilibrium
states with given free energy $F$ inside a cluster $\Gamma$ is given
by
\begin{equation}
d{\cal N}(F)=\exp[-x(q)\beta F_{\Gamma}+x \beta F]dF
\label{dis1}
\end{equation}
Where $F_{\Gamma}$ is the free energy of the cluster.  The cavity
method derives the expression of the free energy and the function
$q(x)$ by simply using the condition that the previous distribution
(\ref{dis1}) must reproduce itself when a new spin $s_0$ is added to a
system of $N$ spins.  If we assume that there is a complexity curve
$\Sigma (f)$ we can apply the cavity method to reproduce the recipe of
Ref. [\onlinecite{MPRL95}].  The only difference with the standard
equilibrium treatment is the starting point, {\it i.e.} the average
number of metastable states with free energies near a given free
energy $F^*=N f^*$, which now reads:
\BEQ
d{\cal N}(F)=\exp \left[N \Sigma (f^*)
- \left. {\partial \Sigma \over \partial f}\right|_{f=f^*}
\hspace*{-.3 cm}F^* 
+ \left. {\partial \Sigma
\over \partial f}\right|_{f=f^*}\hspace*{-.3 cm}F \right]dF
\EEQ
If we  make the formal identification 
\begin{equation}
\beta x= \left.{\partial \Sigma \over \partial f}\right|_{f=f^*}
\end{equation}
the derivation of chapter V of Ref. [\onlinecite{MPV}] can be applied
straightforwardly.  There is a crucial difference, however, between the
two expressions: the first quantity is  $O(1)$ for the
states at  the equilibrium free energy while the second quantity is
$O(e^{\Sigma N})$ for the metastable states.  
Correspondingly, in the equilibrium case the final result coincides
with the true equilibrium free energy, while in the latter case the
result is the free energy of a system in which the states $\alpha$ are weighted
with a non-Boltzmann weight proportional  to $e^{-\beta x
F_{\alpha}}$ where $\beta x=\partial \Sigma(f^*)/\partial f$  is
the Legendre conjugate of $f$. 
 A Legendre transform must therefore be applied to obtain
the complexity, just as in Ref. [\onlinecite{MPRL95}].

In the context of the cavity method the distribution $P(y)$ of the
cavity field $y$ inside a state, required to compute objects like
$\overline{m^k}$, can be identified with the function $P(x,y)$ at the
value of $x$ corresponding to the self-overlap of the states in the
framework of replica computation (see Sec. \ref{s:var}). 
The physical meaning of the functions $P(x,y)$ and $m(x,y)$ at any $x$
can be understood and their evolution operators rederived.  The
function $P(x,y)$ describes the distribution with respect to the
disorder of the average cavity field $y$ of a cluster labelled by
$q(x)$ and is computed taking into account the presence of the spin
$s_0$. The function $m(x,y)$ yields the magnetization, averaged over
the cluster, of the spin $s_0$ when the average cavity field is $y$.
However, the weights assigned to the subclusters to compute the
averages are not the Boltzmann weigths. Indeed, the averages inside a
cluster $\Gamma$ are defined as:

\begin{equation}
\langle {\cal O} \rangle_{\Gamma} ={\sum_{\alpha} e^{-\beta x
F_{\alpha}} \langle {\cal O} \rangle_{\alpha} \over \sum_{\alpha}
e^{-\beta x F_{\alpha}}}
\end{equation} 

Note that the previous average coincides with the Boltzmann averages
only at the level of the configurations  corresponding to $x=1$.
Thus the magnetization inside a state is actually the thermodynamic
magnetization in presence of an average cavity field $y$, {\it i.e.}
$m(1,y)=\tanh(\beta y)$. The peculiarity of these averages is that the
subclusters with lower free energy no longer dominate the average, and
all the subclusters have the same (infinitesimal) weight. As a
consequence, the quantity $\int dy~P(x,y)~m(x,y)^2$, the average of the
overlap inside a cluster, is simply given by $q(x)$, yielding the
variational equation (\ref{q}). Had we weighted the
subcluters with their Boltzmann weight this would not be true
 since few states would have
dominated forcing us to consider diagonal contributions proportional
to the self-overlap.
The cavity method thus yields the evolution operators for $P(x,y)$ and
$m(x,y)$ which coincide with the equations (\ref{eqP}), (\ref{eqm}).

%%%%%%%%%%%%%%%%%%%%%%%%%%%%%%%%%%%%%%%%%%%%%%%%%%%%%%%
\subsection{Overlap Discontinuity and  Replicon Eigenvalue}

In this section we present a simple argument in order to show that the
FRSB quenched solution must be discontinuous at $x_b$ in order to
implement the recipe of Ref. [\onlinecite{MPRL95}] but, at the same
time, this discontinuity causes the replicon eigenvalue
$\Lambda_R=x_{\rm P}$ to be negative.

We start considering the expression of the replicated variational free
energy as a function of $q(x)$ reported in Eq. (\ref{f:Phi}).  This expression
must be extremized with respect to all parameters but $x_b$.
Therefore, as noted in Ref. [\onlinecite{SDJPC84}], at the extremum
the total derivative with respect to $x_b$ coincides with the partial
derivative:

\BEA 
&&{\partial f^{\rm rep} \over \partial x_b} = {\beta \over 4}
\left[q^2(1|x_b)-q^2(x_b^-)\right] 
\\ \nn
 &&\hspace*{ 2 cm}+ {1 \over \beta
x_b^2}\int dy~P(x_b^-,y) 
\\ \nn 
&&\hspace*{.5 cm}\times\Bigl[ \log \int {\cal{D}}z
~p(x_b,y,z)- \left< \log p(x_b,y,z)
\right>\Bigr] 
\EEA
where $p(x_b,y,z)$ is defined is Eq. (\ref{f:p1}).
The above quantity is zero when $\Delta q=0$, therefore we need a
discontinuity of $q(x)$ at $x=x_b$  in order to compute the Legendre
transform, Eq. (\ref{f:SLeg}).

\begin{figure}[th!]
\begin{center}
\includegraphics*[width=.35\textwidth]{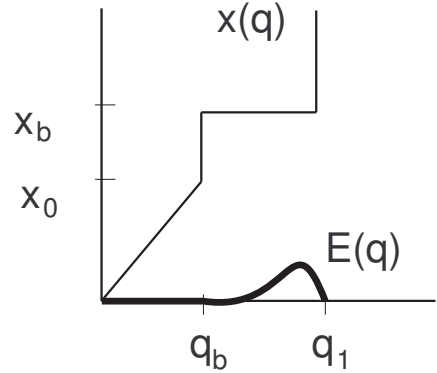}
\caption{Function $x(q)$ and equation $E(q)$. The value $x_0$ is the one
corresponding to the lower end of the $q$-plateau:
$q_b=q(x_0)=q(x_b)$. The maximum overlap value is $q_1=q(1|x_b)$.}
\label{x_q}
\end{center}\end{figure}

In the following we show that the presence of the discontinuity leads
to a negative replicon.  The discontinuity corresponds to a plateau in
the function $x(q)$ starting at $q_b$ and ending at $q_1$. The
function $E(q)$ defined earlier must be equal to zero at $q_b$ and
$q_1$ but not between these points, see figure \ref{x_q}.  We
work under the hipothesys that $\epsilon=x_{\rm c}-x_b$ is small
and that the corresponding function $x(q)$ is similar
to the one  of the equilibrium Parisi
solution ($\epsilon=0$).  
In particular, we assume that
the function $x(q)$ displays a region with $\partial_q x(q)\neq 0$ on the
left of the point $q_b^-$.
This is
precisely the behaviour we have encountered in the analisys of the SK
model  in the preceding sections. 
Our analisys will be valid only at small $\epsilon$, which
corresponds to the region of free energy near the equilibrium free
energy. This region however is quite important since it is natural to
assume that the function $\Sigma (f)$, if it exists, is a continuous
function of $f$ going to zero at the equilibrium free energy.
Furthermore, the outcome of the present analysis is confirmed by
numerical evaluation for which it is verified
in the whole range of $x_b$ values, 
as we will show
in the following.

It is useful to introduce the function $E(q)$:
\begin{equation}
E(q)=q-\int dy~P(q,y)~ m(q,y)^2 
\end{equation}
The variational equations impose that $E(q)=0$ where $\p_q{x}(q)\neq
0$.  Instead, if $x(q)$ is constant, e.g.  for $q_i \leq q\leq q_f$,
we must have $E(q_i)=0=E(q_f)$ at the end points but not in the rest
of the interval.
In other words, if there is a plateau in $x(q)$ [which corresponds to
a dicontinuity in $q(x)$] the equation $q=\int dy~P(q,y)~ m(q,y)^2$
must be verified only at the starting point and at the ending
point. We mention that this is sufficient to show that the FRSB
solution with a discontinuity at $x_b$ considered in this paper must
verify the condition
\BEQ
\int_0^1 dx~q(x)=1-T \ , 
\EEQ
as
the equilibrium solution actually does.

The derivatives of the function $E(q)$ with respect to $q$ can be
 evaluated using the equations for $P(q,y)$ and $m(q,y)$ written above
 and expressed only in terms of derivatives with respect to
 $y$. \cite{S85, CR} After some manipulations we have:

\begin{eqnarray}
&&\p_q E(q) =  1-\int dy~P(q,y) [m'(q,y)]^2 
\label{EE1}
\\
\label{EE2}
&&\p_q^2 {E}(q) =  2\beta x(q)\int dy~P(q,y) [m'(q,y)]^3 
\\
\nn
&&\hspace*{1 cm}-  \int dy~P(q,y) [m'(q,y)']^2 
\\ 
&&\partial_q^3{E}(q) =  2 \beta ~\p_q{x}(q) \int dy~ P(q,y) [m'(q,y)]^3 
\\ \nn
&&\hspace*{.5 cm}-6\beta^2x(q)^2\int dy~P(q,y) [m'(q,y)]^4 
\\
\nn
&&\hspace*{.5 cm}+12 \beta x(q) \int dy~P(q,y)~ m'(q,y)[m''(q,y)]^2
\\
\nn
&&\hspace*{.5 cm}
-\int dy~P(q,y) [m'''(q,y)]^2 
\label{EE3}
\end{eqnarray}
The central point of the argument is the observation that the replicon
eigenvalue can be expressed as the derivative $\p_q {E}(q)$ evaluated
at $q_1$. Indeed at $q_1$ we have $m=\tanh \beta y$ which satisfies
$m'=\beta(1-m^2)$, thus $\p_q {E}(q_1)=x_{\rm P}$.  
This property leads immediately
to the result that each model with a continuous FRSB $q(x)$, such
as the SK model, is marginally stable, see e.g. [\onlinecite{SDJPC84,Som83}]
and Sec. \ref{ss:var1}.

The previous expressions allow to study the behaviour of the
derivatives of $E(q)$ at the point $q_b$ where $x(q)$ is discontinuous
(see fig. \ref{x_q}). Indeed, since the functions $P(q,y)$ and
$m(q,y)$ are continuous at $q_b$ we can relate the derivatives on the
right of $q_b$ to the derivatives on the left of $q_b$, which are
identically zero.  We have, then,

\begin{equation} 
\p_q{E}(q_b^+)=\p_q{E}(q_b^-)\rightarrow \p_q{E}(q_b^+)=0
\end{equation}
\BEA
\p_q^2{E}(q_b^+)&=&2 \beta (x_0+\Delta x)\int dy~P(q,y) [m'(q,y)]^3 
\\
\nn
&&\hspace*{1 cm}- \int dy~P(q,y) [m''(q,y)]^2 
\\
\nn
&=&\p_q^2{E}(q_b^-)+2 \beta \Delta x \int dy~P(q,y) [m'(q,y)]^3
\\
\label{e2}
&=&c_1 \Delta x > 0
\EEA
with $\Delta x= x_b-x_0$.
Analogously we have:
\BEA
\partial_q^3 E(q_b^+)&=&-2~\beta~ \p_q x(q_b^-)\int dy~P(q,y)[m'(q,y)]^3
\nn
\\
&&\hspace*{4 cm}
+O(\Delta x)
\nn
\\
\label{e3}
&=& -c_2+O(\Delta x)<0
\EEA
The last inequalities in (\ref{e2}) and (\ref{e3}) follow from the fact
 that the function $m(q,y)$ entering $\int dy~P(q,y)[m'(q,y)]^3$ is equal to
 $\tanh \beta y$ at first order in $\Delta x$ and therefore at this order
  the integral is strictly positive.  Now the relation
\BEQ
 0=E(q_1)={c_1\over 2} \Delta x \Delta q^2-{c_2\over 6}\Delta
 q^3+O(\Delta q^4)
\EEQ
implies
 \BEQ
\Delta q=(3 c_1/c_2)\Delta x +O(\Delta x^2)
\EEQ 
and
\begin{equation}
\p_q{E}(q_1)=-\frac{3}{2} (\Delta x)^2 {c_1^2 \over c_2}+O(\Delta x^3)<0 \ .
\end{equation}
The previous argument can be generalized to any mean field
spin-glass model with a FRSB phase under the same
hipothesis for the function $q(x)$ at $x_b$ near $x_{c}$, {\it
i.e.} a discontinuity at $x=x_b$ and a region with $\p_q{x}(q)\neq 0$ on
the left side of $q_b=q(x_b)$.  Indeed, in presence of multiple $p$-spin
interaction the only change in $E(q)$ is that we must substitute the
first term $q$ with a function $\p_q{f}(q)$ which depends on the
interaction strength.  
In general, the Plefka parameter $x_{\rm P}$, or equivalently the replicon
$\Lambda_R$, can be expressed as $\p_q^2{f}(q)-\int dy~P(q,y) ~(1-m^2)^2$,
thus it can be again identified  with $\p_q{E}(q_1)$.  The function
$\p_q{f}(q)$ does not depend on $x(q)$ and therefore it does not affect
the difference between derivatives on the left and on the right side
of $q_b$ and the inequalities in Eqs. (\ref{e2}) and (\ref{e3}) still hold,
leading again to a negative replicon.

The argument can also be generalized to soft-spin systems (with
soft-spin distribution normalized to $\langle S^2\rangle$=1):
in general at $q_1$ we have

\begin{equation} 
m(q_1,y)=\frac{\int P(S) S e^{\beta y S}dS}
{\int P(S) e^{\beta y S}dS}
\rightarrow m'=\beta(1-m^2)
\end{equation}
thus the identification of the replicon with the derivative
$\p_q{E}(q)$, computed at $q=q_1=q(1|x_b)$, holds once again.

%%%%%%%%%%%%%%%%%%%%%%%%%%%%%%%%%%%%%%%%%%%%%%%%%%%%%%%
%%%%%%%%%%%%%%%%%%%%%%%%%%%%%%%%%%%%%%%%%%%%%%%%%%%%%%%

\subsection{BRST-SUSY Subextensive Complexity}

Our most important result is that the quenched FRSB BRST-SUSY solution 
here described  must be
rejected on physical ground at any value of free energy apart from the
equilibrium one.

\begin{figure}[t!]
\begin{center}
{
\includegraphics*[width = 0.49\textwidth]{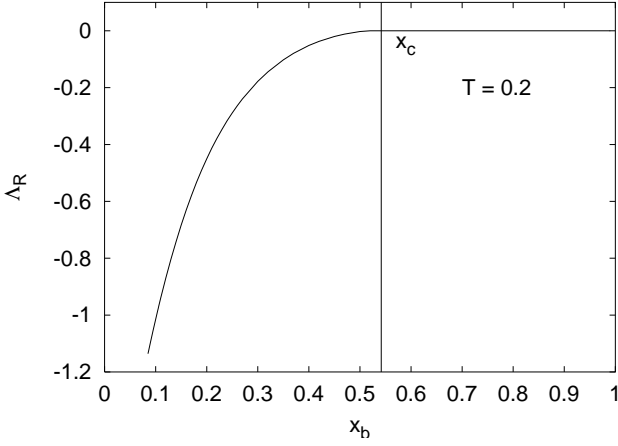}
}
\caption{ Replicon eigenvalue 
$\Lambda_R$ vs. the breaking parameter $x_b$ at $T=0.2$.
Any value of $x_b<x_c$ yields an overlap discontinuity $\Delta q$ causing
a negative $\Lambda_R$.}
\label{fig:R}
\end{center}
\end{figure}

\begin{figure}[ht!]
\begin{center}
{
\includegraphics*[width = 0.49\textwidth]{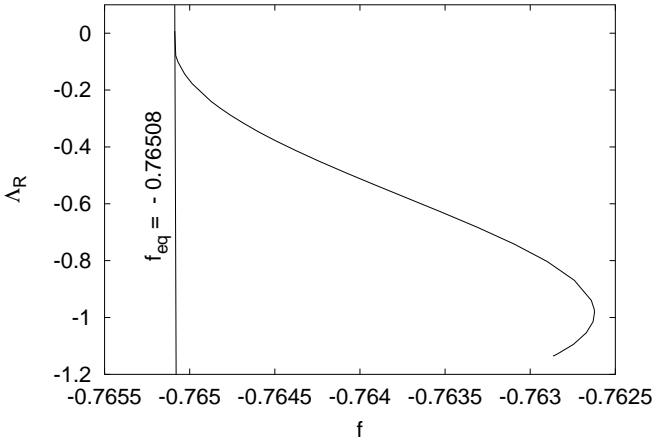}
}
\caption{ Replicon eigenvalue 
$\Lambda_R$ vs. free energy at $T=0.2$. The replicon is negative as soon 
as the free energy landscape is probed above its lowest value.}
\label{fig:R2}
\end{center}
\end{figure}

We have checked 
that the BRST-SUSY solution studied in Sec. \ref{s:BRSTsol}
does not satisfy the Plefka criterion [equation (\ref{xP})], {\it i.e.}
the replicon eigenvalue is negative as soon as $x_b<x_{c}$. 
In figures \ref{fig:R} and \ref{fig:R2}
 we plot the replicon, Eq. (\ref{xp})
versus the breaking parameter $x_b$ and the free energy $f$:
 it turns out to be
always negative for $x_b<x_c$, going to zero from below as $x_b\to
x_c$ and remaining zero for any value of $x_b$ above the
self-consistent point $x_c$.  
Even going beyond the annealed approximation for the SK
model,\cite{CGPM,noian} where the relative 1RSB static phase was not
satisfying Plefka's criterion, the condition $\Lambda_{\rm R} \geq 0$ 
remains violated.  The FRSB BRST-SUSY solution with a non zero $\Delta
q$ is not even marginally stable. In order to have marginal stability
it must be $\Delta q=0$ that means $x_b\geq x_c$, i.e.  the breaking
point cannot be imposed but it has to be determined self-consistently.

For such a marginally stable solution, however, $\Sigma$ is zero,
thus {\em there is no way to implement the Legendre transform approach to
compute any physically well defined complexity}. If BRST-SUSY holds,
there is no exponentially large number of states at any energy level
above $f_{\rm eq}$ in the SK model.

 In terms of free energies, $f_{\rm eq}$ is
equivalent to the lower band-edge $f_0=f(x_c)$, since we are computing
the statics using the right  FRSB scheme.  This means that, provided the
BRST-SUSY complexity actually counts some extensive number of TAP
solutions, they have no physical meaning. As discussed in
Ref. [\onlinecite{noian}] the violation of the Plefka criterion leads
also to a mathematical inconsistency with a given assumption implicit
also in the present solution (i.e. the condition $B=0$, see
Refs. [\onlinecite{BMan,CGPM,BMque,BMY,noian}]). 
As we have previously shown in the present section, a direct computation
indicates that this result can be extended to generic mean-field models
with a continuous FRSB $q(x)$, e.g. the Ising $p$-spin model below the
Gardner transition (see Ref. [\onlinecite{CLRpSP}]).

Since this solution is unphysical we are led to the conclusion
that, holding SUSY, the number of states of the
system is subextensive and the states have all the same free energy
per spin equal to the equilibrium value. Furthermore all the states
verify $x_{\rm P}=0$.

%%%%%%%%%%%%%%%%%%%%%%%%%%%%%%%%%%%%%%%%%%%%%%%%%%%%%%%

\section{Is there a BRST-SUSY-breaking solution?
Quenched FRSB Complexity in the most general scheme of computation.}
\label{s:nonBRST}
In the previuos part of the paper we have been computing the
complexity generalizing the recipe of Monasson \cite{MPRL95} for 1RSB
systems in zero external field to FRSB systems. That means that the
complexity has been built as Legendre transorm of the replicated free
energy, Eq. (\ref{f:Phi}).  In other words, the BRST-SUSY has been assumed
from the beginning.
We look now at the possibility of
computing the quenched complexity
 \BEQ \Sigma_q=\frac{1}{N}{\overline {\log \rho_s(f)} } \EEQ
in a more general formulation,inspired by the Boltzmann microscopic
definition of entropy. It reduces to the one we just employed if
particular supersymmetric relations are satisfied.  The FRSB
expression for the quenched complexity has been presented for the
first time, as far as we know, in Ref. [\onlinecite{BMque}], as
\BEA
&&\Sigma_q=-\lambda~ q - u~f - \Delta (1-q)
-\frac{\Delta^2}{2\beta^2}
\label{Sq}
\\
\nn
&&\hspace*{.5 cm}+\frac{1}{2\beta^2}
\int_0^1\d x\left[\rho^2(x)+\eta^\star(x)\eta(x)\right]
+\phi(0,0,0)
\EEA
where $\phi(0,0,0)$ is the ``equilibrium'', zero local fields
value of the function $\phi(x,h_1,h_2)$ solution of the antiparabolic
three-dimensional differential equation
\BEA
&&\dot\phi=-\frac{\dot\eta}{2}\left[\frac{\p^2\phi}{\p{h_1}^2} 
+x\left(\frac{\p \phi}{\p{h_1}}\right)^2
\right]
\label{Eqh1h2}
\\
\nn
&&-\frac{\dot\eta^\star}{2}\left[\frac{\p^2\phi}{\p{h_2}^2} 
+x\left(\frac{\p \phi}{\p{h_2}}\right)^2
\right]-\dot\rho\left[\frac{\p^2 \phi}{\p{h_1}\p{h_2}}
+x~\frac{\p\phi}{\p{h_1}}\frac{\p \phi}{\p{h_2}}
\right]
\EEA
where the derivative with respect to $x$ is indicated by the dot.
The boundary condition at $x=1$ is
\BEA
&&
\phi(1,h_1,h_2)=\log \int_{-1}^1\hspace*{-.2cm}\d m ~
 e^{{\cal{L}}_{\rm full}(m; u, q, \Delta,
\lambda, \eta(1), \rho(1), \eta^\star(1))}
\nn
\\
&&
\\
&& {\cal L}_{\rm full}(m; u, q, \Delta,
\lambda, \eta(1), \rho(1), \eta^\star(1)) = 
\label{L_full}
\\
\nn
&&
\hspace*{1 cm}-\frac{1}{2}\log\left[ 2 \pi \left(\beta^2 q-\eta(1)\right)
\right]+\log\left(\frac{1}{1-m^2}+B\right)
\\
\nn
&&
\hspace*{2.2 cm}-\frac{\left[\tanh^{-1}m
-\left(\Delta+\rho(1)\right)m +h_1\right]^2}
{2\left(\beta^2 q-\eta(1)\right)}
\\
&&\hspace*{2.2 cm}+\left(\lambda-\frac{\eta^\star(1)}{2}\right)m^2 + u
f(m;q)+h_2m 
\nn
\\ 
&&f(m;q)=-\log 2+\frac{1}{2}\log(1-m^2)
\label{f:fq}
\\ &&\hspace*{ 3 cm}
+\frac{1}{2}m\tanh^{-1}m-\frac{\beta^2}{4}\left(1-q^2\right)
\nn
\EEA 

The order parameters $q, \lambda, \Delta, \eta(x), \eta^\star(x),
\rho(x)$ have to be determined self-consistently by extremising
(\ref{Sq}).  Since we will reformulate everything in the generalized
two-group Ansatz notation we will present the self-consistency
equations later on, once the linear trasformations to that formalism
will be applied.  
We only mention that $\eta(x)$ is proportional to
what was called $q(x)$ up to now and that the BRST-SUSY implies a
dependency of $\Delta$ and $\lambda$ on $q$ and of $\eta^\star$ and
$\rho$ on $\eta$. The number $q$ is the value of all the elements of the
diagonal submatrices of the two-group Ansatz, and it is usually set
equal to $q(1)$ in the unbroken two group scheme.

%%%%%%%%%%%%%%%%%%%%%%%%%%%%%%%%%%%%%%%%%%%%%%%%%%%%%%%%%%%%%

\subsection{BRST relations}
Before going on describing of the two group Ansatz 
we very shortly recall the relations among order parameters
satisfied by systems that are invariant
under BRST-SUSY.

In the BM notation,\cite{BMan,BMque} 
the macroscopic BRST relations, at the level of order parameters,
 are:\cite{noian}
\BEA
&&\Delta = - \frac{\beta^2}{2}u~q
\label{BRSTd}
\\
&&\lambda = \frac{\beta^2}{8}u^2 q
\label{BRSTl}
\EEA
for the annealed approximation and 
\BEA
&&\rho_{ab} = \frac{\beta^2}{2}u~q_{ab}
\label{BRSTr}
\\
&&\eta^\star_{ab} = \frac{\beta^2}{4}u^2 q
\label{BRSTe}
\EEA 
in the quenched RSB case (see  Ref. [\onlinecite{Annibale}],
where a detailed derivation is presented in the notation of Ref. 
[\onlinecite{CGPM}]). 
%%%%%%%%%%%%%%%%%%%%%%%%%%%%%%%%%%%%%%%%%%%%%%%%%%%%%%%
  \subsection{Formulation in the two-group Ansatz}
\label{ss:quenched_2g}
In this section we will first recall the approach of
Ref. [\onlinecite{PP}] in order to reformulate Eqs. (\ref{Sq} -
\ref{L_full}) in the two-group formalism. The motivation to perform
such a transformation is that, from the point of view of the
BRST-SUSY, imposing the BRST relations is equivalent to set equal to
zero all order parameters involved but $q(x)$.

 The two-group Ansatz was introduced by Bray and Moore in
Ref. [\onlinecite{BM2g}] in order to solve the instability problem of
the replica symmetric solution of the SK model.  In
Ref. [\onlinecite{MPRL95}] Monasson showed how the formalism of
Legendre transforms can be applied to mean-field disordered models
through the {\em pinning of real replicas} in a configuration space
extended to $m$ copies of the system.

Parisi and Potters \cite{PP} explained how the BM action can be
obtained through the method of Monasson provided that the symmetry
between real replicas is broken according to a generalized two-group
Ansatz.  Being $n$ the number of replicas introduced to compute the
quenched average 
and $m$ the number of real copies, they analyzed 
\BEQ \lim_{n\to
0}\frac{1}{n}\log{\overline{ Z^{mn}}} =\ext_f\left[{\overline{\log
\rho_s(f)}}- m\beta N f\right]
\label{logZmn}
\EEQ
where $Z^{mn}$ is the partition function of $n\times m$ copies of the
system.  In terms of the replicated matrix-parameter $\hat Q$, its
average is [cfr. Eq. (\ref{a_fq})]
\BEA
&&{\overline{ Z^{mn}}}=
\ext_{\hat Q} \exp\left\{N\left[\frac{\beta^2}{4}
\left(mn-\Tr {\hat Q}^2\right) \right.\right.
\\
\nn
&&\hspace*{1.5 cm}\left.\left.+ \log\sum_{\{\sigma_a^c\}}
\exp\left(\frac{\beta^2}{2}\sum_{ab,cd}{\hat Q_{ab}^{cd}}
\sigma_a^c\sigma_b^d\right)
\right]\right\}
\label{Zmn}
\EEA
where the indexes $a, b = 1,\ldots, n$ while $c, d= 1,\ldots, m$.  The
four index matrix ${\hat Q}_{ab}^{cd}$ can be expressed as the
composition of $n^2$ sub-matrices $\mathbf{Q}_{ab}$ of dimension
$m\times m$ of the form:
\BEQ
\mathbf{Q}_{ab}=\overbrace{\left( \begin{array}{l}Q^+_{ab}  \\
				         Q_{ab}  
			   \end{array}
			   \right.}^{y}
       \overbrace{\left. \begin{array}{r} Q_{ab}  \\
			                  Q^-_{ab}
			   \end{array}
			   \right)}^{m-y}
\label{Q2g}
\EEQ
The matrices $Q_{ab}^{\pm}$ are further parameterized as 
\BEQ
Q_{ab}^{\pm}=Q_{ab}\pm \frac{A_{ab}}{y}+\frac{C_{ab}}{2 y^2}
\EEQ
Furthermore $\hat Q_{aa}^{cc}\equiv 0$.  In Ref. [\onlinecite{PP}] the
last term was denoted by $B_{ab}/y^2$. We write $C_{ab}= 2 B_{ab}$ in
order to obtain more symmetric expressions in the following and to
avoid confusion with the parameter $B$ in Eq. (\ref{L_full}).

Evaluating Eq. (\ref{logZmn}) with this Ansatz, one obtains
a complexity that can be formally connected to the one of
Bray, Moore and Young\cite{BMY} through a given change of variables.

For what concerns the diagonal sub-matrices - corresponding to the 
annealed case - the transformations
bringing from the notation of Ref. [\onlinecite{BMque}] (BM) to the one
of the generalized two-group \cite{PP} (PP) are:
\BEA
\mbox{BM} && \mbox{PP (two-group)}
\nn
\\
q&=&Q
\label{PP_BMq}
\\
\Delta &=&\beta^2\left(A+\frac{m}{2}Q\right)
\label{PP_BMD}
\\
\lambda &=&\frac{\beta^2}{2}\left(C+ m A+\frac{m^2}{4} Q\right)
\label{PP_BMl}
\\
u &=&-m
\label{PP_BMu}
\EEA

Writing  equations (\ref{Sq})-(\ref{f:fq}) with the substitutions 
(\ref{PP_BMq})-(\ref{PP_BMu}), allows for an immediate connection
between the breaking of the matrices structure into two groups
 and  breaking the BRST-SUSY.
 Indeed,
Eqs. (\ref{PP_BMD}) and (\ref{PP_BMl}) transform into the BRST
relations (\ref{BRSTd})-(\ref{BRSTl})
if we set $A=C=0$, i.e. if we do not break the matrix
structure at all ({\em unbroken two-group}).  On the contrary, setting
values of A and C different from zero amounts to break the BRST-SUSY
 and leads to independent values of $q$, $\Delta$ and $\lambda$.

The changes of variables for the off diagonal terms
$\eta_{ab}$, $\eta^\star_{ab}$ and $\rho_{ab}$ are:

\BEA
\mbox{BM} && \mbox{PP (two-group)}\nn
\\
\eta_{ab} &=& \beta^2 Q_{ab}
\label{PP_BMe}
\\
\rho_{ab} &=&- \beta^2\left(A_{ab}+\frac{m}{2} Q_{ab}\right)
\label{PP_BMr}
\\
\eta^\star_{ab}&=& \beta^2\left(C_{ab}+ m~A_{ab} +\frac{m^2}{4}Q_{ab}\right)
\label{PP_BMes}
\EEA
Setting $A_{ab}=C_{ab}=0$ in the two-group construction reproduces 
the generalized BRST 
relations for the replica symmetry breaking case, Eqs. 
(\ref{BRSTr})-(\ref{BRSTe}).

Is there any FRSB solution breaking the BRST-SUSY?  If a complexity
indeed exists the lower band-edge of its dominion in free energy must
coincide with the SK equilibrium free energy $f_{\rm eq}$, computed
making use of the FRSB scheme.  Such a value has been found also by
Bray, Moore and Young: \cite{BMY} starting from the point $ u = u_0$
(such that $\Sigma(u_0) = 0$) a BRST-SUSY solution develops.  Indeed,
their solution was based on an assumption equivalent to the BRST-SUSY
and, therefore, it was not the quenched analogue of the BM solution
(see Sec. \ref{ss:bmy}).

As we just underlined ,
the BRST relations for the order parameters involved in the
computation can be cast in a very straightforward way if the notation
of generalized two-group Ansatz of Ref. [\onlinecite{PP}] is used.

In the Full RSB limit the matrices $Q_{ab}$, $A_{ab}$ and $C_{ab}$
tend respectively to the functions $q(x)$, $a(x)$ and $c(x)$ as $n \to
0$.  In such limit the expression for the quenched complexity
(\ref{Sq}) (with $B=0$) becomes\cite{footnote1}

\BEA
&&\Sigma_q = - \beta uf
-\beta^2(1-Q)\left[A-\frac{u}{2}Q\right]+ \phi(0,0,0)
\nn
\\
&&-\frac{\beta^2}{4}\int_0^1\d x\left[
2 \left(A^2-a^2(x)\right)
-4u\left(A~Q-a(x)q(x)\right)
\right.
\nn
\\
&&\hspace*{2 cm}\left.
+2\left(Q~C-q(x)c(x)\right)
+u^2\left(Q^2-q^2(x)\right)
\right]
\nn
\\
&&
\label{Sq2G}
\EEA
 where $\phi(0,0,0)$ is solution of
\BEA
&&\dot\phi=-\frac{\dot q}{2}
\left[
\frac{\p^2 \phi}{\p y_1^2} +  x\left(\frac{\p \phi}{\p y_1}\right)^2
\right]
-\frac{\dot c}{2}
\left[
\frac{\p^2 \phi}{\p y_2^2} +  x\left(\frac{\p \phi}{\p y_2}\right)^2
\right]
\nn
\\
\label{Eqy1y2}
&&-\dot a 
\left[
\frac{\p^2\phi}{\p y_1\p y_2}
+x~\frac{\p \phi}{\p y_1}\frac{\p \phi}{\p y_2}
 \right]
\EEA
with
boundary condition

\BEA
&&\phi(1,y_1,y_2)=\log \int_{-1}^1 \d m ~e^{{\cal L}_{\rm full}}
\label{b_phi1q}
\\
&&{\cal{L}}_{\rm full}
=\beta u f(m;q)
-\beta m \left(\frac{u}{2} y_1- y_2\right)
\\
&&+\log\left(\frac{1}{1-m^2}\right)
-\frac{1}{2}\log\left[2\pi\beta^2\left(Q-q(1)\right)\right]
\nn
\\
&&
+\frac{\beta^2}{2}
m^2\left[C-c(1)
-u\left(A-a(1)\right)+\frac{u^2}{4}\left(Q-q(1)\right)\right]
\nn
\\
\nn
&&-\frac{1}{2\beta^2\left(Q-q(1)\right)}\Bigl\{\tanh^{-1}m
-\beta y_1
\\
\nn
&&\hspace*{ 1.5 cm}-\beta^2 m \left[\left(A-a(1)\right)
-\frac{u}{2}\left(Q-q(1)\right)
\right]
\Bigr\}^2
\\
&&
\beta f(m;q)=
-\log 2+\frac{1}{2}\log(1-m^2)
\\
&&\hspace*{ 2cm}+\frac{1}{2}m\tanh^{-1}m 
-\frac{\beta^2}{4}\left(1-Q^2\right)
\nn
\EEA
and  where the local fields $h_1$, $h_2$ in Eq. (\ref{Eqh1h2}) 
have been changed in $y_1$, $y_2$
following the linear transformation:
\BEA
&&y_1=-\frac{h_1}{\beta}
\label{y1}
\\
&&y_2=-\frac{u}{2}\frac{h_1}{\beta}+\frac{h_2}{\beta}
\label{y2}
\EEA

Setting 
\BEQ
m\equiv \tanh(\beta y_1+\beta z \sqrt{\Delta q})
\label{def:m_y}
\EEQ 
the above formula (\ref{b_phi1q}) becomes
\BEA 
&&\phi(1,y_1,y_2)=-\frac{\beta^2}{4} u (1-Q^2) 
\\ \nn
&&\hspace*{2.5 cm}+\log \int_{-\infty}^\infty {\cal D}z p(-u,y_1,z)
~e^{{\cal L}_{2}} \\ &&{\cal L}_{2}(z,y_1,y_2)\equiv \beta \tanh(\beta
y_1+\beta z \sqrt{\Delta q}) 
\\ 
\nn 
&&
\hspace*{.5 cm}
\times\left[z\frac{\Delta a}{\sqrt{\Delta q}}+y_2\right.  
\\ 
\nn
&&\hspace*{ 1 cm}\left.+\frac{\beta}{2\Delta q}\tanh(\beta y_1+\beta z
\sqrt{\Delta q}) \left(\Delta q\Delta c-(\Delta a)^2\right)\right]
\label{def:L2}
\EEA
where $p(w,y,z)$ is defined in Eq. 20.

 The self-consistency equations for the order parameters are:
\BEA
&&q(x)=\int dy_1~dy_2~ P(x,y_1,y_2)[\p_{y_2} \phi(x,y_1,y_2)]^2
\label{selfq}
\\
&&a(x)-u~ q(x)=\int dy_1~dy_2~ P(x,y_1,y_2) 
\label{selfa}
\\
\nn
&&\hspace*{3 cm}\p_{y_1} \phi(x,y_1,y_2)
~\p_{y_2}\phi(x,y_1,y_2)
\\
&&c(x)-2 u~ a(x) + u^2 q(x)
\label{selfc}
\\
\nn
&&\hspace*{ 2 cm}=\int dy_1~dy_2~ P(x,y_1,y_2)[\p_{y_1}\phi(x,y_1,y_2)]^2
\EEA

\BEA
&&Q=\left<\left<m^2\right>\right>
\label{saddleQ}\\
&&A=\frac{1}{2Q-q(1)}
\left\{Q a(1) - (Q-q(1)) [1-Q(1+u)] \right\}
\nn
\\
&&\hspace*{.5cm}+\frac{1}{\beta^2[2Q-q(1)]}\left<\left<m\left(
\tanh^{-1}m-\beta y_1 \right)   \right>\right>
\label{saddleA}       
\\
\label{saddleC}
&&C=Q+2A+u (1-2Q+2A)-u^2 Q
\\
\nn
&&\hspace*{1 cm}-2\frac{\Delta a}{\Delta q}\left(A-Q+1-\frac{u}{2}Q
\right)-Q\left(\frac{\Delta a}{\Delta q}\right)^2
\\
\nn
&&\hspace*{1 cm}
-\frac{1}{2\Delta q}\left(1-\frac{1}{\beta^2\Delta q}
\left<\left<
\left(\tanh^{-1}m-\beta y_1\right)^2
\right>\right>\right)
\EEA
where we have made use of Eq. (\ref{saddleQ}) in Eq. (\ref{saddleA}) 
and Eqs. (\ref{saddleQ}), (\ref{saddleA})
to yield Eq. (\ref{saddleC}).
The average $\left<\left<\left(\ldots\right)\right>\right>$ is defined 
as
\BEA
&&\left<\left<O(m,y_1,y_2)\right>\right>\equiv 
\int dy_1~dy_2~ P(1,y_1,y_2)
\\
\nn
&&\hspace*{3 cm}\times\frac{\int_{-1}^1 dm~ 
O(m,y_1,y_2)~e^{\cal{L}_{\rm full}}}
{\int_{-1}^1 dm~e^{\cal{L}_{\rm full}}}
\EEA
or else, using Eq. (\ref{def:m_y}), as

\BEA
&&\left<\left<O(z,y_1,y_2)\right>\right>\equiv 
\int dy_1~dy_2~ P(1,y_1,y_2)
\\
\nn
&&\hspace*{1 cm}\times\frac{\int_{-\infty}^\infty {\cal D} z 
O(z,y_1,y_2)~p(-u,y_1,z) ~e^{{\cal L}_{2}}}
{\int_{-\infty}^\infty {\cal D} z ~p(-u,y_1,z) ~e^{{\cal L}_{2}}}
\EEA
The saddle point equations (\ref{saddleQ})-(\ref{saddleC})
for the elements of the diagonal 
matrices can, then, be written as

\BEA
&&Q=\left<\left<\tanh(\beta~z~\sqrt{\Delta q}+\beta~y_1)^2\right>\right>
\\
&&Q~\Delta a+A~\Delta q=-\Delta q [1-Q(1+u)]
\\
\nn
&&\hspace*{1 cm}+\frac{\sqrt{\Delta q}}{\beta}
\left<\left<
z\tanh(\beta~z~\sqrt{\Delta q}+\beta~y_1)
\right>\right>
\\
&&C=Q+2A+u (1-2Q+2A)-u^2 Q
\\
\nn
&&\hspace*{1 cm}-2\frac{\Delta a}{\Delta q}\left(A-Q+1-\frac{u}{2}Q
\right)-Q\left(\frac{\Delta a}{\Delta q}\right)^2
\\
\nn
&&\hspace*{1 cm}
-\frac{1}{2\Delta q}\left(1-\Delta q\left<\left<z^2\right>\right>\right)
\EEA

The Parisi Eq. (\ref{eqPhi}) and the boundary condition
Eq. (\ref{b_phi1}) can be obtained as a BRST-SUSY
reduction of Eqs. (\ref{Eqy1y2})-(\ref{b_phi1q}), setting 
$A=a(x)=C=c(x)=0$, but leaving $Q\neq q(1)$.
Furthermore, in our notation, $q(1)$ becomes $q(x_b)$
whereas $Q$ is $q(1|x_b)$ as represented in the self-consistency
Eq. (\ref{f:q1}).

If a BRST-SUSY-breaking solution exists it must be a solution of the above
equations with, as {\em necessary} (but not sufficient)
features: $a(x)\neq 0$ and/or $c(x)\neq 0$,
a lower band-edge equal to $f_{\rm eq}$ and a positive
$x_{\rm P}$ decreasing to zero as $f\to f_{\rm eq}$.
 
%%%%%%%%%%%%%%%%%%%%%%%%%%%%%%%%%%%%%%%%%%%%%%%%%%%%%%%
\subsection{An instance of BRST-SUSY solution}
\label{ss:bmy}
  In Ref. [\onlinecite{BMY}] the initial condition of the very same
equation (\ref{eqPhi}) is given by Eq. (14), that we report here for
clarity: 
\BEA
\nn
 && Z(h_1,h_2)= \int_{-1}^1 \frac{dm }{\sqrt{2\pi
[Q_{\rm bm}-\eta(1)]}}\left(\frac{1}{1-m^2}\right) 
\\ &&
\hspace*{1 cm}\exp\Bigl\{-\frac{\left(\tanh^{-1}m-\tilde{\Delta}m
+h_1\right)^2}{2[Q_{\rm bm}-\eta(1)]}
 \nn 
\\
&&\hspace*{2 cm}+\left(\lambda-\frac{1}{2}\eta^\star(1)\right)m^2+u
~f(m;q)+m~h_2\Bigr\} \nn 
\EEA
\BEA 
\nn
&& f(m;q)=-\log
2+\frac{1}{2}\log(1-m^2)
\\ &&\hspace*{2 cm}+\frac{1}{2}m\tanh^{-1}m 
-\frac{\beta^2}{4}\left(1-q_{EA}^2\right)
\nn
\EEA 
The above expression coincides
with Eq. (\ref{L_full}) if we set $B=0$, $Q_{\rm bm}=\beta^2 q$ and ${\tilde
\Delta}=\Delta +\rho(1)$. The symbol $f$ is what we call $\beta f$ in the 
present paper.

Such a function is also identical to our boundary condition
Eq. (\ref{b_phi1}), provided a suitable change of notation is performed.
First of all the $x$-range has to be shifted as
\BEQ
 x\in[0,1] \to x\in[0,x_b]
\label{shift}
\EEQ
so that any parameter computed in $x=1$ will result as computed in $x_b$
in our notation.
Then we identify $u=-x_b$ and 
on the line of transformations (\ref{PP_BMq})-(\ref{PP_BMl}),
(\ref{PP_BMe})-(\ref{PP_BMes})
 we make the following change of variables:

\BEA
&&Q_{\rm bm} =  \beta^2 q(1|x_b)
\hspace*{1.2cm}
\eta(x)=\beta^2 q(x)
\\
&&
\lambda = \frac{x_b^2}{8}\beta^2 q(1|x_b)
\hspace*{1 cm}
\eta^\star(x)=\frac{x_b^2}{4}\beta^2 q(x)
\\
&&
\Delta =\frac{x_b}{2}\beta^2 q(1|x_b)
\hspace*{1cm}
\rho(x)=-\frac{x_b}{2}\beta^2 q(x)
\label{changeD}
\EEA
If $x_b\geq x_c$, $q(1|x_b)=q_{\rm EA}$.
Moreover the local fields are trasformed as in Eqs. 
(\ref{y1})-(\ref{Eqy1y2}), with 
$y_1=y$ and setting $y_2=0$ without loss of generality:
\BEQ
h_1=-\beta y
\hspace*{2cm}
h_2=\frac{x_b}{2}\beta y
\label{h_y}
\EEQ
and the integration variable $m$ (TAP-site-magnetization) is 
changed in $z$ using Eq. (\ref{def:m_y}).

%\BEQ
%m = \tanh\left( \beta y + \beta z \sqrt{\Delta q}\right)
%\label{m_z}
%\EEQ

The function 
$\phi(x,y_1,y_2)$ involved is
 linearly  connected to the solution of Eq. (\ref{eqPhi}) with
boundary condition (\ref{b_phi1}) by the relation

\BEQ
\phi_{\rm BMY}(x,y,0)=\beta x_b
\phi_{\rm Legendre}(x,y)+\frac{\beta^2x_b}{4}(1-q_1)^2
\label{phi2phi}
\EEQ

Exploting Eqs. (\ref{shift})-(\ref{h_y}), (\ref{def:m_y}) and (\ref{phi2phi}),
the complexity becomes equal to the Legendre transform of $f^{\rm rep}$,
given by Eq. (\ref{f:SLeg}).

The actual computation performed by Bray, Moore and Young set, furthermore,
the difference $\Delta q= (Q_{\rm bm}-\eta(1))/\beta^2 = q(1|x_b)-q(x_b^-)$
equal to zero.

%%%%%%%%%%%%%%%%%%%%%%%%%%%%%%%%%%%%%%%%%%%%%%%%%%%%%%%

\subsection{Discussion}
\label{ss:disc}
The resolution of Eq. (\ref{Eqy1y2}) keeping $A$, $C$, $a(x)$ and
$c(x)$ different from zero from the beginning should lead
 to a  BRST-SUSY-breaking complexity. Such a solution would
display some self-consistently determined order parameters, given by
Eqs. (\ref{selfq})-(\ref{saddleC}), breaking the supersymmetry and
allowing for the construction of a quenched complexity, function of
some $x_b$ break parameter and its conjugated $f$.

From what is already known about the annealed complexity and the
property of marginal stability of the SK model at the lowest free
energy values, we expect this solution to fulfill precise
requirements.  First of all there cannot exist states of any kind
below $f_{\rm eq}$, that is the minimal free energy value achievable
by the system at a given temperature. Therefore the lower band edge
$f_0$, that is less than $f_{\rm eq}$ in the annealed approximation,
must coincide with it in the quenched computation.  Moreover the
system is known to be marginally stable, i.e. the replicon, or Plefka
parameter, is zero at $f_{\rm eq}$.

The BRST-SUSY-breaking annealed solution displays a strictly positive
$x_{\rm P}$ for any value of the $f$-conjugated $u$ variable for which
the complexity is positive. 
Numerical evaluations of this parameter, directly counting the number of solutions of the
TAP equations for a $N$ spins system yield different, apparently 
contradictory, results. 
In Ref. [\onlinecite{BMless}] Bray and
Moore performed the direct evaluation of the number of solutions of
TAP equations and in Ref. [\onlinecite{Ple2}] Plefka has carried out
an analogous probe considering modified TAP equations such that only
solutions of the original equations with $x_{\rm P}\geq0$ are
selected.  
In both cases a finite size scaling analysis reveals that the replicon
eigenvalue/Plefka parameter goes to zero as $N^{-1/3}$, thus
indicating an exclusively marginal kind of stability.
On the contrary, a very recent numerical study by Cavagna {\em
et. al}\cite{CGPnum} displays an $x_{\rm P}$ behaviour consistent with
the one obtained for the annealed BRST-SUSY-breaking solution, hinting
 a  strictly positive $x_{\rm P}$  in the whole free energy
support for which $\Sigma > 0$ (see Ref. [\onlinecite{noian}] for the
annelead complexity $x_{\rm P}$).  Even about the free energy density
support numerical findings disagree, Plefka's results hinting for a
vanishing support\cite{Ple2} as $N$ increases, whereas the results of
Ref. [\onlinecite{CGPnum}] show a finite support (even consistent with
the annealed support, at least for $N\leq 80$).

Further analysis would be worthful to better understand the whole picture.
In Ref. [\onlinecite{BMan}]   the solution of the
annealed saddle point equations was shown also to satisfy  the quenched
(replica symmetric) saddle point equations for all $u$ above a
certain, negative, value ${\overline u}$. This includes $u=0$, to
which the largest value of the complexity corresponds.  Over the
$f$-support this is equivalent to say that the annealed solution holds
for $f\in[f({\overline u}),f_{\rm th}]$ but is unstable in the range
$[f_0,f({\overline u})]$. 
Therefore, going to the exact quenched computation  $x_{\rm P}$ should decrease
from $x_{\rm P}(f({\overline u}))>0$ to $x_{\rm P}(f_{\rm eq})=0$ in a well 
defined solution.

The ensemble of TAP-solutions over which the saddle points are computed
is of unknown nature, by this meaning that we cannot know {\em a
priori} if the TAP stationarity points counted by our complexity are
minima or saddles of any order and whether they lead to the correct
linear susceptibility ($x_{\rm P}\geq 0$) or not ($x_{\rm P}<0$).
As a consequence, we are not able to say whether we are counting all
possible solutions (in which case a complexity breaking the
supersymmetric invariance would be mathematical inconsistent) or only
some solutions belonging to a particular subset (in which case the
invariance would not be a necessary requirement anymore).
Recently Apelmeier {\em et. al}\cite{ABM} have shown that the
BRST-SUSY-breaking, else said BM,\cite{BMan} complexity is counting
stationary points of the spin-glass free energy landscape, that are
minima with one flat direction and that lead to $\chi_l=\beta(1-q)$.
The flat direction is bounded to an isolated zero eigenvalue that can
only be found considering subextensive corrections to the complexity
saddle point.  These very same corrections amount to a prefactor of
$\exp N \Sigma^{\rm SP}$ that turns out to be zero at all orders in a
$1/N$ expansion when computed on the BRST-SUSY-breaking saddle
point. This was shown by Kurchan\cite{K91} for the case $u=0$,
corresponding to the maximum value of such complexity, (and to zero
complexity in BRST-SUSY case).  That argument can be extended to any
value of $u$ (or $f$), indicating that the prefactor can be either
exactly zero or exponentially small in $N$.\cite{mmm} Since the Morse
theorem must hold even for the BRST-SUSY-breaking solution the
vanishing prefactor should go like $\exp( - N \Sigma^{\rm SP})$ at
$u=0$.  

The zero prefactor and the isolated eigenvalue are
different ways of expressing the fact that metastable states in the SK
model are of a different nature with respect to those of the $p$-spin
interacting model, with $p>3$.  An isolated eigenvalue has, indeed, been
found also in the spherical $p$-spin model \cite{CLRpSP} but there it
is always positive. This difference is crucial to distinguish between
the dynamical behaviour of spin-glasses and structural glasses
relaxation.  In the case of $p>2$, representing a good mean-field
model for structural glasses, metastable states are proper local
minima (see e.g. [\onlinecite{CS,CGP,CLRpSP}] for the spherical
$p$-spin model or [\onlinecite{CLRpIS}] for Ising $p$-spin) whereas in
the SK model they turn out to be stationary points of different kinds
%display an excape flat direction, 
in the
thermodynamic limit.

%%%%%%%%%%%%%%%%%%%%%%%%%%%%%%%%%%%%%%%%%%%%%%%%%%%%%%%

%%%%%%%%%%%%%%%%%%%%%%%%%%%%%%%%%%%%%%%%%%%%%%%%%%%%%%%

\section{Conclusions}
The main motivation for our work has been the needing for a clearer 
understanding of the  two different approaches that have been
presented for the  behavior of the complexity in the SK
model up to now:  the first one was originally introduced by Bray and Moore
\cite{BMan} 
and 
the second one  initially proposed by Parisi and Potters \cite{PP} and
explicitely carried out (in a somehow different conceptual framework)
quite recently  in Ref. [\onlinecite{CGPM}].

In Ref. [\onlinecite{PP}] it was shown that 
the complexity could be obtained by calculating the
partition function of $m$ distinct real replicas of the system
\cite{MPRL95} and provided the connection with the previous BM
formalism by means of a generalization of the {\em two-group
Ansatz}. \cite{BM2g}
 In that framework they assumed an 
 'unbroken' two-group Ansatz in order to perform explicit
computation.  This has been finally performed 
 exploiting the knowledge of BRST 
SUSY \cite{BRST} (formally equivalent to the unbroken two-group)
\cite{noian} and
leading  to a supersymmetric complexity different the BM one.\cite{CGPM}

 The difference with the BM complexity has turned  out not to be at
the functional level, but simply the new solution is a second saddle
point of the same functional.\cite{noian}  This means that there is
just one BRST-SUSY functional, the BM complexity functional,\cite{BMan,PP}
 displaying (at
least) two saddle points, one satisfying the BRST-SUSY\cite{CGPM} and
the other one breaking it\cite{BMan}.  Both solutions identify an extensive
complexity, computing the number of solutions of the
Thouless-Anderson-Palmer (TAP) equations \cite{TAP} in the annealed
approximation.  In Ref. [\onlinecite{noian}] the authors analyzed the
two annealed solutions and their problems reaching the conclusion that both of
them had to be discarded as candidate complexities of the SK model.

In this paper we have faced the quenched calculation and the analysis
of the properties of the complexity of the Sherrington-Kirkpatrick
model making use of the Full Replica Symmetry Breaking Ansatz.  Such
complexity can be seen as the Legendre transform of the replicated
free energy potential. A detailed description of the behavior of the
complexity is presented in Sec. \ref{s:BRSTsol}.  
Results are obtained both by numerical extremization of the replica
free energy in the FRSB limit at a given temperature below the
critical one ($T_s$) and by anaytical expansion around the critical
temperature.  The complexity we find is the quenched analogue of the
annealed complexity studied in Ref. [\onlinecite{CGPM}]. As such, it
satisfies BRST-SUSY invariance. Even though going to the exact
quenched computation cures some deficiency intrinsic in the annealed
approximation (e.g. the lower band-edge now coincides with the
equilibrium free energy) other problems arise or old inconsitencies
are not overcome.  
The most important outcome is that any attempt to build a BRST-SUSY
complexity using the Legendre transform approach at finite temperature
leads to a {\em negative replicon everywhere in the complexity
support} apart from the lower band-edge/equilibrium value $f_{\rm
eq}$, where the replicon is zero and replica stability is marginal.
This means that the solutions of the mean field TAP equations that our
observable is counting are not physically relevant. More specifically
they do not express a local linear susceptibility equal to
$\beta(1-q_{\rm EA})$. This is related, in the replica language, to an
instability against overlap fluctuations.  Moreover, this leads to an
even more serious problem, since a negative $x_{\rm P}$ is
mathematically inconsitent with the $B=0$ solution (see
Sec. \ref{s:nonBRST}) of the saddle points equations for complexity,
as shown in the Appendix of Ref. [\onlinecite{noian}].

We are, then, brought to the conclusion that a complexity satisying BRST-SUSY,
such as, e.g. the one of $p$-spin models ($p>2$),
cannot yield any extensive contribution in the SK model. 
A subextensive complexity was yield by the
analysis performed by van Mourik and Coolen\cite{vMCJPA01} in a
dynamical reformulation of Parisi solution in terms of a multiple
(infinite) hierarchy of decoupled time-scales, in which a small
fraction of slower spins plays the role of an effective quenched
disorder for the faster spins. Since in their construction the
quantity of 'slow' spins turns out to be subextensive with respect to
the quantity of 'fast' spins, the entropy of the slowest dynamic
variable, i.e. the comlexity, is zero.
This means, though, that a ``supersymmetric assumption'' is hidden somewhere
in their procedure.

We eventually  discussed the possibility of computing a BRST-SUSY-breaking
quenched  complexity, by means of the tools presented in Sec. \ref{s:nonBRST}.
 Even though the nature of the metastable states counted by such
 complexity would be different from the one of structural glass metastable
 states, and conceptually less intuitive, an extensive,
 thermodynamically stable, complexity can, in principle, still be computed.

%%%%%%%%%%%%%%%%%%%%%%%%%%%%%%%%%%%%%%%%%%%%%%%%%%%%%%%
\acknowledgments 
{We thank A. Annibale, A. Cavagna, T.C.C. Coolen,
I. Giardina, T. Plefka, E. Trevigne for interesting exchange of
opinions and information.}

%%%%%%%%%%%%%%%%%%%%%%%%%%%%%%%%%%%%%%%%%%%%%%%%%%%%%%%

\section*{Appendix}

In this appendix we show the expansion of some observables in the
'reduced' temperature $\tau\equiv T_s-T$, where $T_s=1$ is the
critical temperature of the SK model. The following expansion
are strongly asymptotics at higher orders,\cite{CR2} but, 
as already anticipated
in Sec. \ref{ss:pade}, Pad\'e resummation allows for converging
results.
 
The maximum complexity as a function of the reduced temperature is:

\begin{widetext}
\BEA
\Sigma^{\rm max}(\tau) &=& {\frac{{{\tau }^6}}{81}} -
 {\frac{2\,{{\tau }^7}}{81}} 
+  {\frac{187\,{{\tau }^8}}{1215}} 
 - {\frac{9938\,{{\tau }^9}}{10935}}  
 + {\frac{2313541\,{{\tau }^{10}}}{382725}} 
- \frac{10350722\,{{\tau }^{11}}}{229635} 
+  \frac{19122138382\,{{\tau }^{12}}}{51667875} 
\\
\nn
&-& \frac{57172333507\,{{\tau }^{13}}}{17222625} 
+ \frac{128525178819148\,{{\tau }^{14}}}{3978426375}
 - {\frac{182162380772635931\,{{\tau }^{15}}}
    {537087560625}} +  {\frac{133461584583918966554\,
      {{\tau }^{16}}}{34910691440625}} 
\\
\nn
&-& 
  {\frac{964957153958428975243\,{{\tau }^{17}}}
    {20946414864375}} 
+ \frac{591101435671536495956948\,
      {{\tau }^{18}}}{999715254890625}+O(\tau^{19})
\EEA

The difference between the maximum value of quenched
and annealed  complexities is given by

\BEA
\Sigma^{\rm max}(\tau)-\Sigma^{\rm max}_a(\tau)&=&{\frac{{{\tau }^8}}{243}} 
- {\frac{2\,{{\tau }^9}}{135}} +   {\frac{3394\,{{\tau }^{10}}}{54675}}
-   {\frac{1381\,{{\tau }^{11}}}{4725}}  
+  {\frac{9052283\,{{\tau }^{12}}}{7381125}} 
-  {\frac{788437\,{{\tau }^{13}}}{273375}} 
\\
\nn
&-&   {\frac{6757091354\,{{\tau }^{14}}}{217005075}} 
+  {\frac{84666457147717\,{{\tau }^{15}}}{107417512125}} - 
  {\frac{136466739459370883\,{{\tau }^{16}}}
    {10741751212500}} 
\\
\nn
&+&   {\frac{744513309305871199\,{{\tau }^{17}}}
    {3989793307500}} - {\frac{50994683697175207152217\,
      {{\tau }^{18}}}{18851773377937500}} +O\left(\tau^{19}\right)
\EEA

 The difference between the threshold  free energies yield by
quenched and annealed computation is
\BEA 
f_{\rm th}(\tau)-f^a_{\rm th}(\tau)&=&
{\frac{{{\tau }^7}}{162}} - {\frac{13\,{{\tau }^8}}{810}} +
{\frac{2711\,{{\tau }^9}}{54675}} - {\frac{70474\,{{\tau}^{10}}}{382725}} 
+ {\frac{1701148\,{{\tau }^{11}}}{5740875}} 
+{\frac{453457547\,{{\tau }^{12}}}{82668600}} 
\\
\nn
&-&{\frac{269738513263\,{{\tau }^{13}}}{2411167500}} +
{\frac{779531487166381\,{{\tau }^{14}}}{477411165000}} -
{\frac{27180847504347788\,{{\tau }^{15}}}{1220653546875}} 
\\
\nn
&&+{\frac{211583950213407904679\,{{\tau }^{16}}}{698213828812500}} -
{\frac{13746420017310240661979\,{{\tau }^{17}}}{3258331201125000}}
+O\left(\tau^{18}\right)
\EEA
\end{widetext}

%%%%%%%%%%%%%%%%%%%%%%%%%%%%%%%%%%%%%%%%%%%%%%%%%%%%%%%%%%%%%%

\end{document}